\documentclass[aps,prb,reprint,twocolumn,superscriptaddress]{revtex4-1}

\usepackage{xurl}
\usepackage{float}
\usepackage[utf8]{inputenc}
\usepackage{amssymb}
\usepackage{amsthm}
\usepackage{graphicx}
\usepackage{xcolor}
\usepackage{hyperref}
\usepackage{makecell}
\usepackage{booktabs}
\DeclareUnicodeCharacter{2212}{-}
\usepackage{hyperref}
\hypersetup{
    colorlinks=true,
    linkcolor=blue,
    filecolor=blue,
    urlcolor=blue,
    citecolor=blue,
    pdfborder={0 0 0}
}
\begin{document}

\title{Impact of Data Bias on Machine Learning for Crystal Compound Synthesizability Predictions}

\author{Ali Davariashtiyani}
\affiliation{Department of Civil, Materials, and Environmental Engineering, University of Illinois Chicago, Chicago, IL, USA}
\author{Busheng Wang}
\affiliation{Department of Chemistry, University at Buffalo State University of New York, Buffalo, NY, USA}
\author{Samad Hajinazar}
\affiliation{Department of Chemistry, University at Buffalo State University of New York, Buffalo, NY, USA}
\author{Eva Zurek}
\affiliation{Department of Chemistry, University at Buffalo State University of New York, Buffalo, NY, USA}
\author{Sara Kadkhodaei}
\email{sarakad@uic.edu}
\affiliation{Department of Civil, Materials, and Environmental Engineering, University of Illinois Chicago, Chicago, IL, USA}

\date{\today}

\begin{abstract}
Machine learning models are susceptible to being misled by biases in training data that emphasize incidental correlations over the intended learning task. In this study, we demonstrate the impact of data bias on the performance of a machine learning model designed to predict the synthesizability likelihood of crystal compounds. The model performs a binary classification on labeled crystal samples. Despite using the same architecture for the machine learning model, we showcase how the model's learning and prediction behavior differs once trained on distinct data. We use two data sets for illustration: a mixed-source data set that integrates experimental and computational crystal samples and a single-source data set consisting of data exclusively from one computational database. We present simple procedures to detect data bias and to evaluate its effect on the model's performance and generalization. This study reveals how inconsistent, unbalanced data can propagate bias, undermining real-world applicability even for advanced machine learning techniques. 
\end{abstract}
\maketitle

\section{Introduction}\label{sec:intro}
The increasing availability of large materials datasets has facilitated the rapid growth of machine learning for materials discovery and design \cite{ml_importance_Butler, HimanenGeurtsFosterEtAl2019, Pablo:2019aa,Tian2021,Oganov2019,Umehara2019,Jablonka2020,HIMANEN2020106949,Morgan2020,Hart2021,Gong2021,Zhong2022,Damewood2023,Xu2023}. However, the impact of data selection strategies, data quality, or data bias has been less explored or not clearly demonstrated, even for the most advanced machine learning models, although these factors can significantly impact model performance, generalizability, and reliability \cite{Agrawal2016,databias_liang2022}.
In this study, we specifically investigate the effect of \textit{data heterogeneity} on machine learning models. For demonstration, we use a machine learning model designed for predicting crystalline compounds' synthesis feasibility (or synthesizability), previously developed by our group \cite{Davariashtiyani_2021_SYN}. Our machine learning model for synthesizability prediction is an ideal platform because data selection for such a model is inherently challenging. A portion of the data must represent already synthesized materials (synthesizable) from crystal structure datasets, while another portion must be artificially generated to represent hypothetical crystals unlikely to be synthesized or formed (unsynthesizable). This challenge has led to arbitrary data selection in the literature, resulting in different strategies. For example, studies in Ref.\cite{syn_Frey2019,syn_Jang2020,syn_Antoniuk2023,syn_Gleaves2023} assume that unsynthesizable examples are not available and thus predicts 2D or crystalline materials'  synthesizability based on a positive and unlabeled (PU) classification model \cite{ml_PU} on data from a single dataset. In contrast, Ref.\cite{Davariashtiyani_2021_SYN} predicts crystalline materials' synthesizability based on a binary classifier on labeled data from different sources. Here, we examine the generalization performance of our synthesizability model for two data selection approaches: whether data is coming from different sources (heterogeneous data) or from an identical source  (homogeneous data). 

%
%
Inherent data bias is a common challenge in machine learning \cite{databias_liang2022}, stemming from sampling bias, data collection bias, domain bias, or labeling bias. The bias in data usually leads to spurious correlations and biases picked up by the model, regardless of the choice of the machine learning algorithm \cite{databias_liang2022}. In the field of data-driven materials research, some studies have explored the influence of data on the performance of machine learning models \cite{data_bias_Kumagai2022,data_bias_zhang2023entropy,data_bias_Li2023,Zhang2018,De_Breuck_2021}. For example, Kumagai \textit{et al.} demonstrated that data bias influences the error and reliability of predictions made by a machine learning model \cite{data_bias_Kumagai2022}. Using the experimental property data from Starrydata2 database as a demonstration platform,
they defined an applicability domain - a material space to which the machine learning model can be applied based on its similarity to known materials used for training. They revealed that model predictions are more reliable within the applicability domain. Within the applicability domain, prediction accuracy remains high, while outside the defined applicability domain, accuracy significantly decreases. They also observed that the prediction error decreases within the applicability domain as the number of neighboring known materials increases. 
%
In another study, Zhang \textit{et al.} \cite{data_bias_zhang2023entropy} introduced an information entropy-based metric to measure data bias and developed an entropy-targeted active learning (ET-AL) framework to mitigate it. They utilized the ET-AL framework to guide new data acquisition in density functional theory (DFT)-generated materials databases (such as OQMD~\cite{db_oqmd_1, db_oqmd_2} and JARVIS~\cite{db_jarvis}) by addressing the imbalanced coverage of formation energy among different crystal systems (i.e., structure-stability bias). This approach aimed to enhance the diversity of underrepresented crystal systems and, as a result, improved the performance of machine learning models \cite{data_bias_zhang2023entropy}.
%
%
In a separate study, Li \textit{et al.}~\cite{data_bias_Li2023} investigated the impact of distribution shifts (or domain shifts) on the performance of machine learning models, offering solutions for diagnosing, anticipating, and addressing this challenge. Data distribution can undergo significant shifts, even between different versions of an actively expanding database, owing to changes in preferences or focus over time.  For instance, they illustrated a significant decline in the performance of a formation energy prediction model trained on Materials Project 2018 (e.g., ALIGNN-MP18\cite{models_str_ALIGNN}) when applied to new compounds in the Materials Project 2021 database\cite{db_mp}, with prediction errors ranging from 23 to 160 times larger than those observed when the model is tested on Materials Project 2018. By utilizing a uniform manifold approximation and projection (UMAP) approach as a measure of similarity to cluster the data, they observed that test samples with low prediction errors tend to reside within clusters covered by the training data (i.e., similar to the training data). Conversely, the majority of poorly predicted test samples form a distinct, isolated cluster separate from the rest of the data. To enhance prediction robustness and generalizability of the machine learning model, they introduced both UMAP-guided and query-by-committee data acquisition strategies.
 Other studies noted the impact of data bias on machine learning model performance \cite{davariashtiyani_2023_FE,Jiang2021}. For example, our group observed that the bias in the DFT materials database, specifically the imbalanced distribution of negative and positive formation energies in the Materials Project, results in diminished prediction performance for larger, positive-value ranges of formation energy and is the likely cause for the progressive increase of error from metallic to ionic materials \cite{davariashtiyani_2023_FE}. 
As a demonstration platform for exploring data bias, we utilize a machine learning model developed by our group to predict the synthesizability of crystal compounds \cite{Davariashtiyani_2021_SYN}, albeit with significant modifications that enhance its performance (as detailed in the Methods section). Therefore, we present a concise review of machine-learning methods applied to materials synthesis, categorizing these studies into two groups: 
First group are studies that focus on developing machine learning models for predicting the synthesis feasibility (i.e., synthesizability) of given products or crystal compounds \cite{Polymorphism_Jones2017,Davariashtiyani_2021_SYN, syn_Frey2019, syn_Jang2020, syn_Zhu2023, syn_Antoniuk2023, syn_Gleaves2023}. The demonstrative model used in this study belongs to this group. These models typically involve learning correlations between chemical/structural patterns in existing crystal compounds and a score of synthesizability (or synthesizability likelihood). As mentioned earlier, data selection and labeling are difficult in such models due to the absence of unsynthesizable crystal compounds. The second group of studies aims to develop models for predicting synthesis routes or reactions (e.g., solid-state, sol–gel, or solution - hydrothermal, precipitation), synthesis procedures, synthesis conditions (e.g., temperatures, times), or synthesis precursors or reactants \cite{syn_Raccuglia2016_failed, syn_kim_2017_params_routes, syn_Huo2019_procedures, syn_Kononova2019_recipes, syn_kim_2020_precursor, syn_karpovich2021inorganic_rout_condition, syn_Wang2022_procedures, syn_Huo_2022_condition, syn_Karpovich_2023_condition, syn_McDermott_2023_procedures}. These studies encompass a range of approaches, from data-driven learning of materials synthesis information using natural language processing of existing scientific literature \cite{syn_kim_2017_params_routes, syn_Huo2019_procedures, syn_Kononova2019_recipes, syn_kim_2020_precursor, syn_karpovich2021inorganic_rout_condition, syn_Wang2022_procedures, syn_Huo_2022_condition} to the development of graph-based networks based on thermodynamic and kinetic data (i.e., physics-informed) \cite{syn_Aykol2019_network, syn_Aykol2021_pareto, syn_McDermott2021_network, syn_McDermott_2023_procedures}. The latter approach is employed for predicting chemical reaction pathways in solid-state materials synthesis. 
In this study, we analyze the performance of a crystal compound synthesizability model trained on two sets of data through a comparative modeling experiment. In the first experiment, data is collected from two distinct sources, creating a mixed-source or heterogeneous dataset. Synthesizable crystal compounds are sourced from the Crystallography Open Database (COD) \cite{Grazulis2009}, while unsynthesizable samples are computationally generated using the Crystal Structure Prototype Database (CSPD) \cite{db_cspd}. Details of the generation of unsynthesizable samples are provided in the Methods section. In the second experiment, data for both classes is collected from a single dataset, specifically the DFT-generated Materials Project database \cite{db_mp}. We refer to this dataset as a single-source or homogeneous dataset. Although we use the exact crystal compound samples in both experiments, in the latter, the atomic configurations of the reported crystal compounds in Materials Project are relaxed through DFT geometry and cell optimization. Therefore, crystal structure data from Materials Project corresponds to a zero-pressure and no-applied-stress condition. In both experiments, we employ the same machine learning architecture - a convolutional neural network (CNN) for processing sparse voxel image representations of crystals connected to a binary classifier. Further details of the machine learning model are provided in the Methods section. We compare the predictive performance and generalizability of both models and investigate the underlying reasons for performance differences.
\section{Methods}\label{sec:methods}
\begin{figure*}[h]
    \centering
    \includegraphics[width=\textwidth]{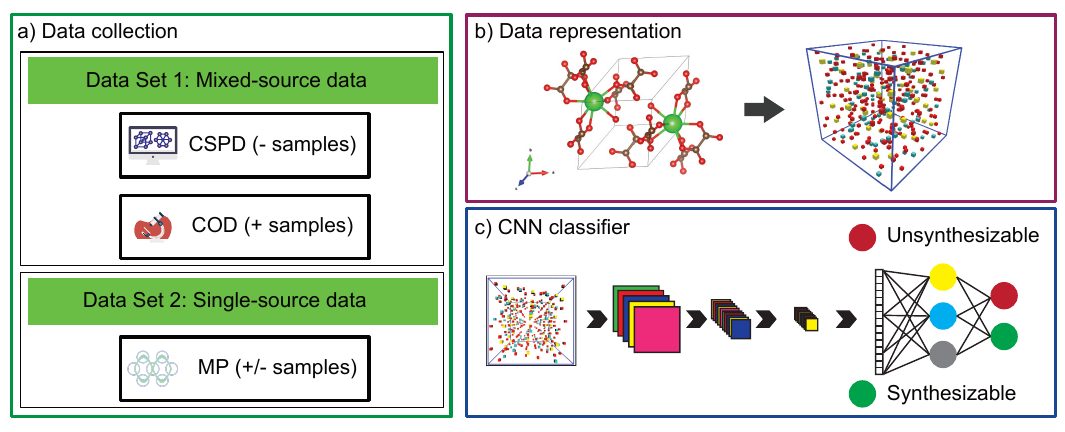}
    \caption[Overall]{The overall framework of the synthesizability likelihood prediction model. a) Data collection for the two distinct data sets used in this study. For data set 1 or mixed-source data, crystal samples for the synthesizable class are obtained from the Crystallographic Open Database (COD) and crystal samples for the synthesizable class are generated using the Crystal Structure Prototype Database (CSPD). For data set 2 or single-source data, crystal samples for both classes are collected from the Materials Project database. b) The crystal information files (CIFs) are converted into color-coded voxel images, which are used as the inputs for the convolutional encoder. c) The convolution encoder followed by a multi-layer perceptron (MLP) binary classifier trained on labeld data, referred to as the CNN classifier.}
    \label{fig:overall}
\end{figure*}

\textbf{Crystal Synthesizability Model}: 
Our crystal synthesizability model consists of a convolutional neural network (CNN) connected to a neural network classifier. The interconnect CNN and neural network perform feature learning and classification tasks, respectively, on  labeled crystal structure data. This model operates on sparse voxel images of crystals, as developed by our group \cite{Davariashtiyani_2021_SYN,davariashtiyani_2023_FE}. Our voxel image representation creates a 3D visual depiction of the crystal structure color-coded by the identities of its constituting chemical elements (see Ref. \cite{davariashtiyani_2023_FE} for details). The voxel images are created in a cubic box with a 50 $\AA${} edge. The atoms in the crystal unit cell are repeated to fill the cube. Then the cube is partitioned into a voxel grid of size 128$\times$128$\times$128. The voxel images are RGB color-coded with the three channels representing atomic number, group number, and valence number, respectively. The CNN encodes the hidden structural and chemical patterns of crystal compounds by processing the sparse voxel images into a low-dimension set of latent features. The latent features are input into the neural network classifier (a binary classifier). The model is trained on labeled crystal images, representing two classes of synthesizable and unsynthesizable crystals. 
The architecture of the synthesizability model is shown in Fig. \ref{fig:overall} and is detailed in our previous work \cite{Davariashtiyani_2021_SYN}. The CNN consists of a sequence of three blocks, each consisting of a convolution, an activation, and a pooling layer. The CNN architecture flattens the output and propels it through a 3-layer dense neural network with a (13,13,13,1) node architecture. This model is operated on our most recent framework for sparse voxel image representation of crystalline materials as detailed in Ref. \cite{davariashtiyani_2023_FE}. Compared to our previous study.~\cite{Davariashtiyani_2021_SYN}, the synthesizability model used in this work adopts a more advanced voxel image representation (as detailed in Ref.~\cite{davariashtiyani_2023_FE}). Additionally, the presented synthesizability model employs augmentation of rotated crystal samples (data augmentation) during training and an ensemble averaging technique during prediction to improve model’s consistency and rotational invariance. Details of the data augmentation and ensemble averaging are provided in Ref. \cite{davariashtiyani_2023_FE} with a brief discussion in Appendix \ref{sec:ensemble}. 
\\

\noindent\textbf{Labeling Approach}
The binary classification is performed on two classes of crystalline materials: synthesizable versus unsynthesizable crystals, the latter being the hypothetical crystalline materials that are unlikely to be synthesized. The positive or synthesizable class comprises experimentally synthesized crystal compounds readily available in crystal databases (e.g., ICSD (Inorganic Crystal Structure Database) \cite{db_icsd}, COD \cite{Grazulis2009}). On the other hand, the negative or unsynthesizable class must represent crystal compounds that are unlikely to form or be experimentally synthesized, at least based on the existing scope of synthesis techniques and conditions available to us. Therefore it lacks a
dedicated repository, which makes the collection of data for this class challenging. 
Other models for synthesizability in previous studies \cite{syn_Frey2019, syn_Jang2020, syn_Gleaves2023} use positive-unlabeled (PU) learning as they assume the absence of explicit negative class samples. In contrast, we use an \textit{a priori} labeling approach by carefully selecting negative class samples. Our strategy involves identifying the top 0.1\% (108 compositions or chemical formulas) of well-studied crystal compounds in the materials science literature from 1922 to 2021 using natural language processing (see details in Ref. \cite{Davariashtiyani_2021_SYN}). For these compositions, we assign a negative label to crystal structure polymorphs not found in existing experimental databases. Crystal samples come from either a single source or multiple sources, as explained below.

In contrast to our \textit{a priori} labeling approach, alternative synthesizability models employ a semi-supervised learning strategy, incorporating both labeled and unlabeled data (PU learning). This involves learning characteristics of negative samples through a data-driven machine learning technique known as pseudo-labeling. Notable examples include the transductive bagging scheme utilized in Ref. \cite{syn_Frey2019, syn_Jang2020, syn_Antoniuk2023}, and the dynamic entropy-based pseudo-labeling within a teacher-student dual neural network \cite{syn_Gleaves2023}. In PU learning, the model is trained to learn characteristics associated with positive samples by distinguishing them from the `average' characteristics of unlabeled data, according to a similarity or distance measure between unlabeled and positive samples. In contrast, our synthesizability model learns distinguishing characteristics of positive and negative samples from explicit examples of each class. Therefore, it can be considered a more supervised approach that incorporates a level of human-based physical interpretation of the negative samples.
%

Both the semi-supervised learning approach and the explicit labeling approach introduce biases into the data and subsequent learning processes (i.e, labeling bias).
In our labeling approach in this work, inherent bias arises due to the limited chemical distribution of explicit negative samples, encompassing only 108 chemical compositions as examples of the negative class. We opt against expanding the chemical distribution of negative samples by selecting a larger percentile of top-studied compositions in the literature. This choice is motivated by the need to maintain confidence in designating negative examples, avoiding potential mislabeling.
On the other hand, the PU learning approach may introduce other biases. For instance, any unlabeled sample representing a hypothetical crystal compound can contribute to patterns correlated with a negative sample through a weighted average scheme based on similarity measures. However, many of these unlabeled samples collected from computational databases are examples of undiscovered or unexplored synthesizable polymorphs. Therefore, the PU approach has the potential to introduce implicit mislabeling biases to the learning process. Identifying and addressing implicit biases associated with negative samples are beyond the scope of this study. Instead, our focus is on examining the biases introduced through the collection of examples for the positive and negative classes, whether from a single database or multiple databases. 
\\

\noindent\textbf{Mixed-Source Data Collection}
The mixed-source dataset is compiled from two distinct databases. Positive examples, representing synthesizable materials, are sourced from the experimental Crystallographic Open Database (COD) ~\cite{Grazulis2009, Quiros2018, Merkys2016, Grazulis2015, Grazulis2012, Downs2003}, while negative samples, indicating unsynthesizable materials, are generated using the Crystal Structure Prototype Database (CSPD)\cite{db_cspd}. The procedure for generating negative samples is as follows: Initially, we select the topmost studied compositions in the literature, resulting in 108 unique compositions (details provided in Ref.~\cite{Davariashtiyani_2021_SYN}). These chemical compositions are then input into the CSPD toolkit to generate all possible hypothetical crystal polymorphs. For a given chemical formula, CSPD selects known crystal prototypes from its database to act as templates. In this process, elemental sites are substituted with the desired chemical elements, and the lattice parameters are adjusted to match a target volume. Finally, the inter-atomic distances of the generated structures undergo validation. Details of utilizing CSPD for generating crystal samples is given in Appendix \ref{sec:cspd}. From the structures generated by CSPD, we filter out those having identical crystal structures in COD, designating them as belonging to the positive class. Following this procedure, we generated 597 negative samples. For the positive class, we limit the collection to 2,960 crystal samples from COD to maintain a balanced sample size between positive and negative classes. The positive samples encompass all crystal polymorphs for the selected 108 chemical compositions available in COD (367 crystals), with the remaining samples randomly chosen from other compositions (2633 crystals). Some of these crystals are filtered out due to image resolution constraints (see Ref.~\cite{Davariashtiyani_2021_SYN} for details), resulting in 2,960 positive crystal samples passable to the model. The data including positive and negative samples is split into training (60\%), validation (20\%), and test (20\%) sets. 
\\

\noindent\textbf{Single-Source Data Collection}
The single-source dataset is exclusively compiled from crystal structures in Materials Project (MP v2022.10.28)  \cite{db_mp}, which is a DFT database for crystal compounds. To ensure consistency between the two datasets for the comparative analysis in this study, we reference the same chemical compositions constituting the mixed-source dataset. For the compositions in the positive class, any MP crystal structure entry that matches the composition and has an ICSD tag is collected as a positive sample. MP entries that match the mixed-source dataset's negative class compositions without an ICSD tag are identified as negative samples. Following this approach, a total of 1068 positive and 930 negative crystal structures are collected from MP. We split the data into training (60\%), validation (20\%), and test (20\%) sets. 

\section{Results}
\subsection{Data Bias Detection}
To identify bias within a dataset, one needs to define metrics for measuring such bias. In this study, we employ density and mean atomic mass as two metrics to detect potential biases in both single-source and mixed-source datasets. These features, chosen to represent fundamental characteristics of crystal samples, are selected with the awareness that they are not expected to exhibit a direct correlation with synthesizability. Noticeable differences in the distributions of these basic features between the two classes indicate the presence of data bias that could adversely impact model performance. Distribution discrepancies across datasets have been used in other studies to reveal potential data bias. For instance, Kumagai \textit{et al.} \cite{data_bias_Kumagai2022} compared the distributions of average atomic masses and average electronegativities across different databases, including ICSD, Materials Project, Magnetic Materials, and Starrydata2. The observed differences in distributions were considered as a potential source of data bias.

\begin{figure*}[h]
    \centering
    \includegraphics[width=\textwidth]{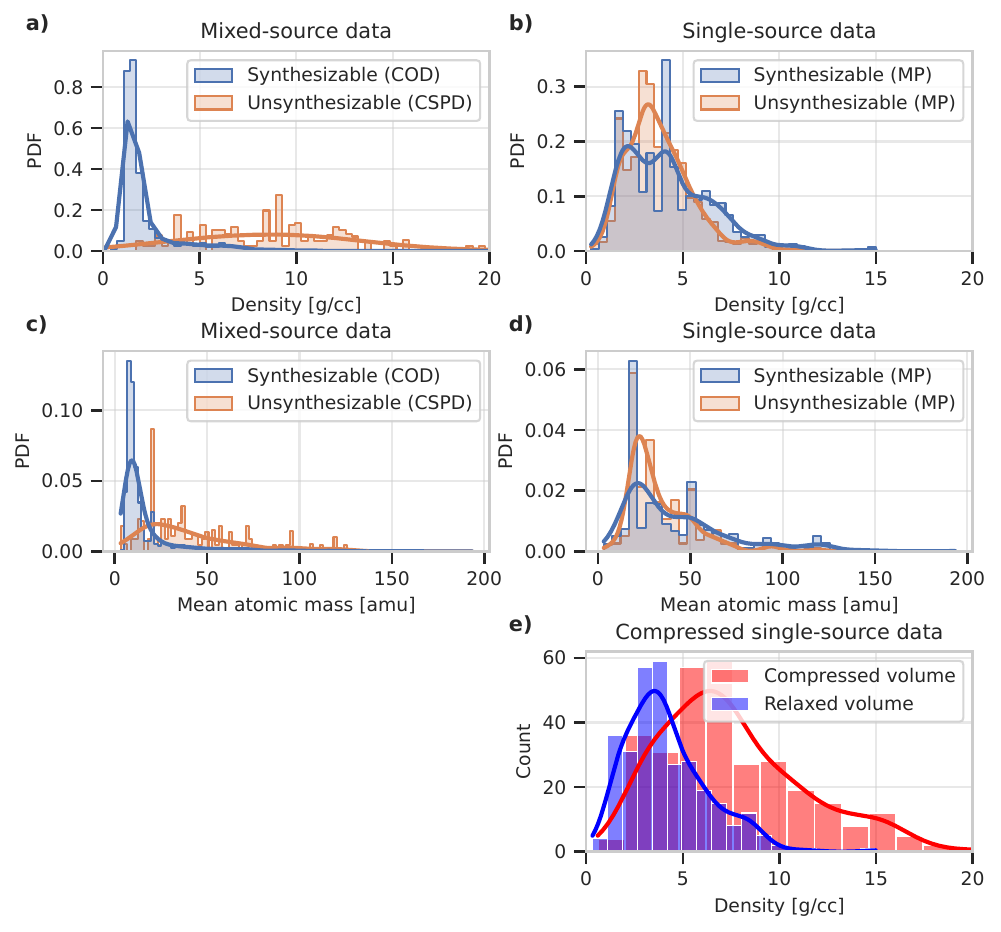}
    \caption[Data set comparison]{Density and mean atomic mass distributions of crystal samples across different classes and datasets.  (a), (b) Density distributions of crystal samples (synthesizable versus unsynthesizable) in the mixed-source and single source datasets, respectively. (c), (d) Mean atomic mass distribution of synthesizable versus unsynthesizable crystal structures in the mixed-source and single-source datasets, respectively.  e) Density distributions of crystal samples of the single-source test set, shown for their original (relaxed) volume and compressed volumes by 55\%. Both synthesizable and unsynthesizable samples are included in the test set.}
    \label{fig1new:comparison}
\end{figure*}

Fig. \ref{fig1new:comparison}(a,b) compares the density distributions between synthesizable and unsynthesizable samples for both the single-source and mixed-source datasets. Density is calculated as the total atomic mass divided by the crystal structure volume. As shown in Fig. \ref{fig1new:comparison}(a), the density distribution of synthesizable samples is significantly different from the unsynthesizable class in the mixed-source data. Unsynthesizable samples demonstrate a substantially wider density range with a flat peak and a higher average density. Synthesizable samples exhibit a relatively sharp peak around 2 g/cc, while unsynthesizable sample densities widely spread from 2 to 15 g/cc. The higher densities observed in unsynthesizable samples are likely a result of the low-fidelity procedure used to generate hypothetical crystal samples in CSPD, especially when assigning a target volume to the hypothetical crystal. This contrasts with a more accurate albeit computationally intense approach, such as relaxing the crystal volume based on DFT forces (as is done for any hypothetical crystal in the MP database). CSPD constructs a new chemical arrangement of atoms on a crystal prototype skeleton based on simple norms (e.g., composition, symmetry, and configuration) and similarity to available crystal prototypes in its database. As a result of this high-throughput procedure, the target volumes assigned by CSPD to the unsynthesizable samples are systematically smaller than the equilibrium (or relaxed) volume. It is important to note that the majority of experimentally synthesized crystals are associated with ambient pressure conditions. For example, our query on the 2023 version of the ICSD database shows that 208954 crystal enteries are associated with pressures below 1 MP (i.e., ambient presusre) while only 748 crystals entries are for pressure above 1 MP. The distribution of ICSD crystal samples over pressure (above 1 MPa) is shown in Fig. \ref{fig:pressure} in Appendix \ref{sec:icsd_dist}. While the pressure information in ICSD signifies the external conditions during synthesis, it is reasonable to assume that the resulting synthesized crystal reaches equilibrium with its external environment and thus corresponds to the external pressure condition. The systematically larger density of the unsynthesizable samples compared to synthesizable ones indicates a source of bias in the mixed-source data. The illustration of data bias in Fig. \ref{fig1new:comparison}(a) and previous studies \cite{data_bias_Kumagai2022} underscores the need for caution when collecting data from distinct databases, as is the case in this study with mixed-source data. Such data become susceptible to inherent biases measured based on different basic characteristics. 
The single-source data indicates similar density distributions between the synthesizable and unsynthesizable samples, as shown in Fig. \ref{fig1new:comparison}(b). The density distributions for positive and negative samples are both centered around 3 g/cc, although the positive class shows a bimodal distribution with peaks slightly below and above the center. Unlike the mixed-source data, the single-source data does not exhibit any systematic density shift between the synthesizable and unsynthesizable samples.

Apart from density, we compare the mean atomic mass distributions of the two classes in the single-source and mixed-source datasets, as shown in Fig. \ref{fig1new:comparison} (c,d). We observe noticeable differences in the mean atomic mass distributions between the synthesizable and unsynthesizable classes in the mixed-source data. However, this difference is not as significant as the density distributions. As illustrated in Fig. \ref{fig1new:comparison}(c),  in the mixed-source data, the synthesizable samples cluster around lower mean atomic masses compared to the unsynthesizable ones. In contrast, the mean atomic mass is more evenly distributed between the synthesizable and unsynthesizable samples in the single-source data (see Fig. \ref{fig1new:comparison}(d)). The discrepancy in mean atomic mass distributions in the mixed-source data likely arises from the more diverse chemical compositions accessible in the CSPD database compared to the MP database. In both the single-source and mixed-source data, we draw unsynthesizable samples from the exact same 108 unique chemical formulas. However, the high-throughput nature of the CSPD approach for generating crystal structures results in a more diverse set of crystal structures covering most of the 108 compositions, thereby exhibiting a diverse distribution among different mean atomic masses. On the other hand, for a computationally expensive DFT database such as MP, crystal samples associated with these 108 chemical formulas are less diverse, leading to a sharper peak in the mean atomic mass distribution for negative samples in the single-source data (see Fig.\ref{fig1new:comparison}(d)). Comparing Fig.\ref{fig1new:comparison}(c) and (d)  indicates a more balanced coverage of mean atomic mass between the positive and negative classes in the single-source data, while the negative class shows a larger average mean atomic mass in the mixed-source data. This distribution discrepancy or imbalance is a result of bias introduced in the mixed-source data rather than any realistic correlation between synthesizability and mean atomic mass.

The baseline feature comparisons in this section reveal clear evidence of data bias between the synthesizable and unsynthesizable samples in the mixed-source data. The balanced distribution of density and mean atomic mass between the two classes in the single-source data highlights the benefits of compiling data from a single source to minimize potential biases. If data collection from distinct sources is necessary, then practices should be considered to make the data from different sources consistent or mitigate the bias they introduce into the model. In the next section, we demonstrate how our machine learning model is susceptible to learning the incidental associations introduced by the bias between classes rather than capturing the intended structural and chemical signatures of synthesizability likelihood.
\subsection{Impact of Data Bias on Model Performance}
\begin{figure*}[h]
    \centering
    \includegraphics[width=\textwidth]{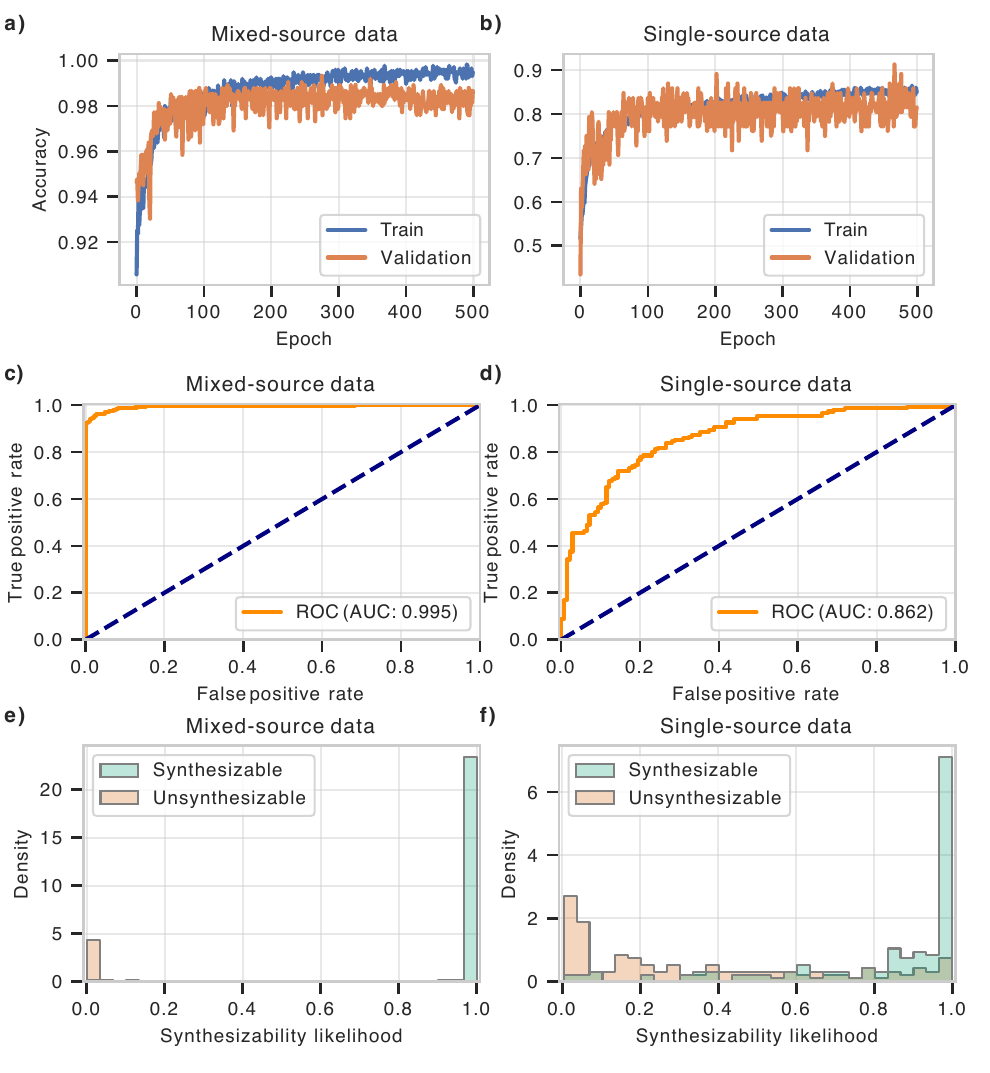}
    \caption[Training and evaluation]{The synthesizability model's training history and performance using mixed-source and single-source data. (a), (b) learning curves of the synthesizability model on the mixed-source and single-source data, respectively. The learning curves show the classification accuracy over epochs, for both training and validation sets. (c), (d) Receiver operating characteristic (ROC) curves of the synthesizability model on the mixed-source and single-source data, respectively. The area under the curve (AUC) is shown for each model. (e), (f) Normalized distribution of synthesizability likelihood of crystals in the test sets of the mixed-source and single-source data, respectively.}
    \label{fig2new:history_evaluation}
\end{figure*}

\noindent\textbf{Learning Performance}
Fig. \ref{fig2new:history_evaluation}(a,b) compares the learning curves of the synthesizability model trained on both the single-source and mixed-source datasets. It illustrates the model’s accuracy on the training and validation sets over 500 epochs. Interestingly, the model exhibits different learning behaviors on the two datasets. Training on the mixed-source data results in rapid convergence, reaching a 90\% accuracy on both the training and validation sets after the first epoch. The model exhibits minimal incremental learning over the subsequent epochs, and eventually reaches a near-perfect accuracy within less than 50 epochs. This rapid convergence suggests that the model distinguishes between synthesizable and unsynthesizable samples too easily, likely biased by the larger density of the unsynthesizable samples, rather than learning inherent synthesizability features. An apparent gap in prediction accuracy between the training and validation sets also suggests overfitting, although accuracy is surprisingly high on both sets. In contrast, when trained on the single-source data, the model exhibits a much slower convergence, starting with a modest 52\% accuracy in the first epoch, which is as good as random guessing, before stabilizing at 80\% after 100 epochs. Notably, the model trained on the single-source data does not show signs of over- or under-fitting, with both training and validation accuracy reaching 80\%. During training, we apply a random rotation to each crystal image at each epoch to promote approximate rotation invariance (see details in Ref.\cite{davariashtiyani_2023_FE}). 

The disparity between the learning curves highlights that the classification success on the mixed-source data may be spurious and is likely influenced by inherent bias between class samples. In other words, we hypothesize that the bias in the training data has misled the model to learn features that are not truly indicative of the learning objective. The observations in this section elucidate that while the model's architecture is maintained, the nature and systematic differences within training examples can significantly influence model's learning behavior.\\

\noindent\textbf{Evaluation on Test Set}
We assess the performance of synthesizability models trained on both single-source and mixed-source data using their respective test sets. The synthesizability likelihood for each sample in the test set is averaged over an ensemble of 100 randomly rotated images of the crystal. Additional details on the choice of the ensemble size are provided in Appendix \ref{sec:ensemble}. For evaluating the classification performance on the test data, we employ metrics such as the area under the receiver operating characteristic curve (AUC-ROC), accuracy, precision, recall, and specificity (a classification threshold of 0.5 is utilized). Fig.\ref{fig2new:history_evaluation}(c,d) compares the prediction performance of the model trained on mixed-source versus single-source data. In Fig.\ref{fig2new:history_evaluation}(c) the ROC curve for the mixed-source synthesizability model is illustrated, showing a high AUC value of 0.995. The mixed-source model exhibits high values for accuracy (0.967), precision (0.990), recall (0.971), and specificity (0.95), indicating the model's near-perfect performance in positive predictions as well as maintaining low false-negative predictions. The synthesizability likelihood distribution of test samples in the mixed-source model, as shown in Fig.\ref{fig2new:history_evaluation}(e), indicates that the model can distinctly differentiate between samples of the two classes. Almost all negative and positive samples are classified with synthesizability likelihoods of 0 and 1, respectively. This clear separation implies that the classification task is straightforward for the mixed-source model. 

In Fig.\ref{fig2new:history_evaluation}(d), the ROC curve for the single-source model is presented. In comparison to the mixed-source model, the single-source model demonstrates an overall lower prediction performance, with an AUC of 0.862. The single-source model's accuracy is 0.778 (compared to mixed-source accuracy of 0.967). Recall remains reasonably high at 0.859 (compared to mixed-source accuracy of 0.971), indicating the single-source model's ability to detect positive samples with a low number of false negatives. The precision and specificity of 0.749 and 0.69 suggests a relatively higher number of false positive and negative predictions, respectively. Fig.\ref{fig2new:history_evaluation}(f) illustrates the distribution of the single-source model’s synthesizability likelihood, depicting the false positive and negative predictions.

The evaluation of the two models’ performance on their respective test data reveals very similar trends to their performance on their training and validation sets. While both models are considered good classifiers according to key metrics, the mixed-source model demonstrates notably superior performance. In the next sections, we demonstrate that this superior performance is biased and thereby not reliable.

The evaluation exercise in this section demonstrates typical procedures widely used to assess machine learning model performances, specifically by evaluating the model's predictions on the test set. However, we illustrate that such evaluation procedures are not suitable for detecting spurious or biased learning by the model or the detrimental effects of data bias on the model's performance.\\ 

\begin{figure*}[h]
    \centering
    \includegraphics[width=\textwidth]{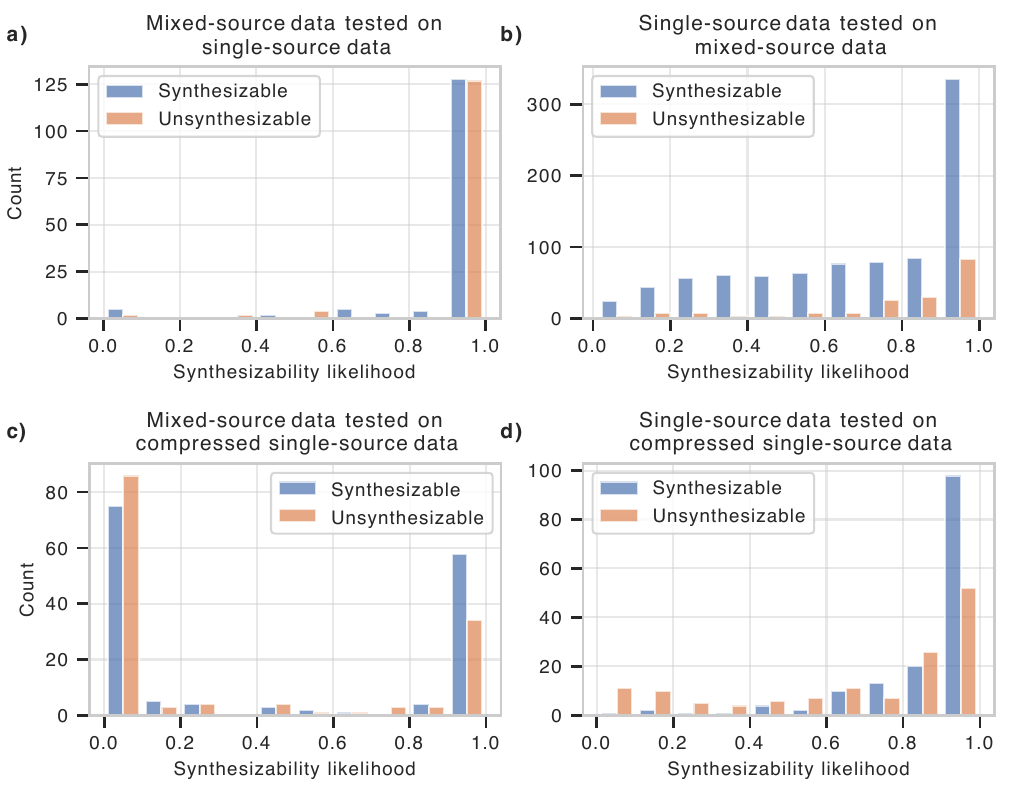}
    \caption[Evaluating on different sets]{
    Evaluation of synthesizability model performance outside its test set under two scenarios: cross-evaluation on swapped test sets and assessment on compressed crystal data. Cross-evaluation assesses the performance of two models (mixed-source and single-source models) on each other's test sets. a) Distribution of synthesizability likelihood for crystals in the single-source test set predicted by the mixed-source model, with accuracy, precision, recall, and specificity of 0.51, 0.51, 0.94, 0.04, respectively. b) Distribution of synthesizability likelihood for crystals in the mixed-source test set predicted by the single-source model, with accuracy, precision, recall, and specificity of 0.63, 0.81, 0.73, 0.15, respectively. c) Distribution of synthesizability likelihood for compressed crystals in the single-source test set predicted by the mixed-source model, with accuracy, precision, recall, and specificity of 0.56, 0.61, 0.43, 0.7, respectively. d) Distribution of synthesizability likelihood for compressed crystals in the single-source test set predicted by the single-source model, with accuracy, precision, recall, and specificity of 0.61, 0.58, 0.94, 0.23, respectively.}
    \label{fig3new:evaluation_comparison}
\end{figure*}

\noindent\textbf{Cross-Evaluation on Swapped Test Sets}
To extend beyond test set evaluation, we assess the classification metrics of both single-source and mixed-source models on swapped test sets. Specifically, we evaluate the mixed-source synthesizability model (trained on data from COD and CSPD) on the test set derived from single-source data (collected from MP). While the mixed-source synthesizability model performs near perfectly on its original test set, it encounters challenges in differentiating between unsynthesizable and synthesizable samples in the single-source test data. There is a significant performance drop in this scenario, with accuracy, precision, recall, and specificity dropping from 0.967 to 0.51, 0.990 to 0.51, 
0.971 to 0.94, and 0.95 to 0.04, respectively. In Fig. \ref{fig3new:evaluation_comparison}(a), the distribution of synthesizability likelihood for single-source test samples is depicted using the mixed-source model. As observed in Fig. \ref{fig3new:evaluation_comparison}(a), the combination of a high number of false positives and a close-to-zero number of false negatives (high recall and very low specificity) implies that the model mistakenly identifies all crystal samples as synthesizable or positive. This observation can be explained by the fact that the mixed-source model tends to categorize any crystal sample within a density range of approximately 1.5-4 g/cc as synthesizable, a pattern learned from the bias in its training data (refer to Fig.\ref{fig1new:comparison}(a)). This density range corresponds to the majority of samples in the single-source data (refer to Fig.\ref{fig1new:comparison}(b)), including the test samples in this exercise. In essence, the mixed-source model has predominantly learned variations in crystal density rather than chemical and structural synthesizability attributes. This observation supports our hypothesis that the mixed-source model has captured a spurious correlation between density and synthesizability likelihood, derived from the bias in its training data. As demonstrated in the previous section, this spurious correlation cannot be detected through a standard test set evaluation procedure. 

We also assess the performance of the single-source synthesizability model (trained on data exclusively from MP) on the test set derived from mixed-source data. As anticipated, the model's performance decreases compared to when evaluated on its original test set; however, the performance drop is less severe than that of the mixed-source model. The accuracy, precision, recall, and specificity of the single-source model are 0.63, 0.81, 0.73, 0.15, respectively, when tested on the mixed-source test data. Fig. \ref{fig3new:evaluation_comparison}(b) depicts the distribution of synthesizability likelihood of mixed-source test samples using the single-source model. An interesting observation emerges: the recall is relatively high (0.73), indicating a low number of false negatives. In other words, the single-source model correctly predicts most of the synthesizable crystals as synthesizable. On the other hand, the specificity is low (0.15), indicating that many of the unsynthesizable samples are incorrectly predicted as synthesizable. The primary reason for the notable difference between recall and specificity is that the positive samples in the mixed-source data set share similarities in their density and mean atomic mass with both the positive and negative samples observed and learned by the single-source model  (see Fig.\ref{fig1new:comparison}). In simpler terms, the positive samples in the test set fall within the applicability domain of the single-source model. In contrast, the negative samples in the mixed-source dataset differ in their density and mean atomic mass from both the positive and negative samples learned by the single-source model, placing them outside the applicability domain of the model. This discrepancy explains the less accurate prediction of unsynthesizable samples. Kumagai \textit{et al.} demonstrated that the prediction performance of machine learning models significantly decreases outside their applicability domain \cite{data_bias_Kumagai2022}. 

Another important observation is that the prediction performance of the single-source model on mixed-source negative test samples is better than the mixed-source model's performance on single-source negative test samples. Specifically, the single source model's specificity on the mixed-source data is still larger than the mixed-source model's on the single source data (0.15 vs. 0.04), indicating that the single-source model generalizes better outside of its applicability domain. Although the number of mispredictions are high (attributed to test data being outside the applicability domain), the single-source model predicts a broader range of synthesizability likelihoods on negative samples compared to the mixed-source model tested on single-source data (compare Fig. \ref{fig3new:evaluation_comparison}(a) and (b)). This observation suggests that the single-source model does not recognize density as the primary correlated attribute to synthesizability. This further confirms our hypothesis that a model trained on heterogeneous or mixed-source data is susceptible to capturing spurious correlations.\\

\noindent\textbf{Crystal Compression Test}
To further test our hypothesis that the mixed-source model primarily bases its decisions on density, we conduct a compression experiment on crystal samples in the single-source test data (exclusively from MP). We then assess the performance of both the mixed-source and single-source models on the compressed crystal structures. For this experiment, we apply a 55\% isotropic volume reduction to each crystal sample in the single-source test set. Figure \ref{fig1new:comparison}(e) illustrates the density distribution of the test set samples before and after the volume reduction. As shown in Fig. \ref{fig3new:evaluation_comparison}(c), the mixed-source model predicts many of the samples as unsynthesizable. This result strongly supports our hypothesis because the same model could not recognize any of the same examples as unsynthesizable when uncompressed (compare Fig. \ref{fig3new:evaluation_comparison}(a) and (c)). In other words, the mixed-source model predictions largely vary with the density variation of the crystal sample, indicating that the model has established a strong correlation between crystal structure density and synthesizability. Additionally, on the compressed data, the mixed-source model mispredicts many of the positive samples as unsynthesizable when compressed (compare the blue bars in Fig. \ref{fig3new:evaluation_comparison}(a) and (c)). This showcases that the chemical or structural features underlying synthesizability of crystal compounds are not learned by the mixed-source model. On the other hand, the predictions of the single-source model on compressed single-source data did not change significantly compared to its predictions on the mixed-source data (compare Fig. \ref{fig3new:evaluation_comparison}(b) and (d)). This implies that the single-source model does not identify any strong correlation between crystal density and synthesizability likelihood. The performance of the single-source model drops on the compressed crystal test samples compared to the uncompressed test samples, with its accuracy reducing from 0.778 to 0.61 (compare Fig. \ref{fig2new:history_evaluation}(f) and Fig. \ref{fig3new:evaluation_comparison}(d)). However, this performance drop is much less significant compared to the mixed-source model, especially on the positive samples, showcasing the better generalization of the single-source model. As mentioned earlier, this performance drop is related to the use of the model outside its applicability domain. Even when utilized on the compressed crystal samples, the single-source model's performance on positive samples remains very good (recall of 0.94), indicating that the model has in fact learned the underlying chemical or structural features of synthesizability.  

The compression test exercise in this section introduces a simple evaluation procedure for assessing how the performance of a model is influenced by the bias in data. In this study, we observed a clear bias in the density of crystal samples in the mixed-source data. Therefore, creating test data with systematic variation in their density through the compression test provides a reasonable platform to detect the propagated bias in the models’ learning and performance.

\section{Conclusion}
This study compares the learning behavior and prediction performance of a machine learning model trained on two different data sets. The model performs a classification task to differentiate between samples of synthesizable versus unsynthesizable crystal compounds. To set a data set for the binary classification, we follow two data collection procedures: The mixed-source data set consists of data from separate computational and experimental crystal structure databases while the single-source data set consists of data from a single computational source. We detect a clear bias in the density and mean atomic mass distribution between samples of the two classes in the mixed-source data. Our results indicate that the mixed-source model produces biased and unreliable predictions, although all standard classification metrics suggest it is a near-perfect classifier. On the other hand, the single-source model shows a less accurate yet more reliable prediction performance. This study presents simple evaluation procedures beyond standard evaluation practices in the literature to measure the effect of data inconsistency on the model’s prediction performance.

In conclusion, we underscore the potential for obscure inconsistency in data resulting from data collection across multiple sources. As demonstrated in this study, the collection of crystal structure samples for binary classification can introduce inconsistencies, particularly in their density distribution. This form of data bias is easily overlooked, given that it remains undetectable through standard evaluation procedures commonly employed in typical machine learning studies for materials prediction or discovery. Hence, it becomes imperative to systematically compare data across various properties, such as density, chemical composition, and structural distribution, in order to identify any potential imbalances or sources of bias before initiating model training.

While demonstrating the detrimental effect of data bias on machine learning models, this study does not lay out pathways for mitigating data bias in the model’s learning and performance, although we showcase the superior performance and generalization of the a model trained on homogeneous data. However, the path forward should involve careful data selection methods to promote diverse, unbiased datasets. Mitigating data bias will be crucial for achieving reliable, useful machine learning predictions in materials science. Data bias detection and mitigation studies in the literature \cite{data_bias_Kumagai2022,data_bias_zhang2023entropy,data_bias_Li2023,Zhang2018,De_Breuck_2021}, while few, provide the roadmap for future studies in machine learning applied to materials science.

\clearpage
\appendix
\section{Rotational ensemble averaging}\label{sec:ensemble}
We employ an ensemble averaging method for predicting the synthesizability likelihood of crystal compounds. Once a crystal sample is input into the trained model, a ensemble of $N$ randomly rotated instances of the sample is generated and the synthesizability likelihood prediction is averaged over the ensemble, where $N$ is a adjustable parameter of the model denoting the ensemble size. The ensemble averaging methods improves the prediction accuracy and robustness of our model, as shown in our earlier study for formation energy prediction \cite{davariashtiyani_2023_FE}, and as illustrated in the following for synthesizability prediction.

Table \ref{tab3:rotation_exercise} illustrates the single-source model's prediction consistency and stability on its test set in terms of the ensemble size. As the ensemble size increases, the number of inconsistent predictions dramatically decreases. The prediction on a crystal sample is considered consistent if the predicted labels for all the rotational instances of the crystal sample are identical, and inconsistent otherwise. A negative label is assigned if the predicted synthesizability likelihood is below 0.5 and a positive label is assigned otherwise (i.e., a classification threshold of 0.5 is utilized). There exist 288 distinct crystal samples in the single-source model test set. As shown in Table \ref{tab3:rotation_exercise}, for an ensemble of size 1, 123 out of the 288 samples have inconsistent predictions, resulting in consistent prediction percentage of around 56\%. A rapid monotonic increase in prediction consistency is observed as the ensemble size increases, leading to a consistency percentage of 90\% and above 96\% for ensemble sizes of 30 and 100, respectively. Another metric we use to measure the consistency is the standard deviation of synthesizability likelihood predictions for a given crystal sample. The standard deviation is measured over $N$ rotational instances of the crystal compound. Table \ref{tab3:rotation_exercise} shows the average standard deviation of synthesizability likelihood of the inconsistent crystal samples in the test set in terms of the ensemble size. A descending trend in the standard deviation as the ensemble size grows is demonstrative of the model's enhanced confidence in its predictions. This suggests that larger ensembles translate to more consistent and confident results. For example, for an ensemble of size 50, the average standard deviation over the 11 inconsistent crystal samples stands at a mere 0.011. Notably, the small standard deviation indicates that the inconsistent predictions corresponds to crystal samples located close to the decision boundary, where the predictions should be close to 0.5 with small oscillations of around 0.011 that render their labels positive or negative.

We select an ensemble size of 100 for our model prediction. For an ensemble size of 100, the model exhibits a high consistency percentage of 95.7\% on its test samples, while simultaneously maintaining a low average standard deviation of 0.015 over the inconsistent samples. Beyond ensemble 100, the incremental improvements in both the 
consistency percentage and average standard deviation are relatively minor. Given these results, ensemble 100 emerges as an optimal balance between performance and computational efficiency. It offers high consistency and confidence in predictions without the necessity for considerably larger ensemble sizes that would demand more computational resources without a sufficient improvement in model performance. 

\begin{table*}[h]
    \centering
    \resizebox{0.8\textwidth}{!}{
    \centering
    \begin{tabular}{|c|c|c|c|}
\hline
\thead{Ensemble size} & \thead{Number of Inconsistent Predictions} & \thead{Consistent Prediction Percentage} &  \thead{Average Standard Deviation of \\ Inconsistent Predictions}  \\
\hline
\hline
1 & 123 & 56.4\% & 0.109 \\

2 & 79 & 72.0\% & 0.085 \\
5 & 45 & 84.0\% & 0.055 \\
10 & 39 & 86.2\% & 0.039 \\
20 & 33 & 88.3\% & 0.030 \\
30 & 27 & 90.4\% & 0.024 \\
50 & 18 & 93.6\% & 0.020 \\
75 & 17 & 94.0\% & 0.017 \\
100 & 12 & 95.7\% & 0.015 \\
150 & 11 & 96.1\% & 0.011 \\
200 & 9 & 96.8\% & 0.010 \\
\hline
\end{tabular}
    }
    \caption[Agreement between Predicted Labels]{The agreement between predicted labels of the 288 samples of the single-source model test set in terms of the ensemble size used for each sample prediction. A sample is deemed consistent if all predicted labels are identical, and inconsistent otherwise. The consistent prediction percentage column represents the fraction of samples with consistent predicted labels among the 288 samples. The standard deviation of an inconsistent sample is measured among all the copies in the ensemble. The average standard deviation over inconsistent samples is presented in the last column.}
    \label{tab3:rotation_exercise}
\end{table*}
\section{CSPD Atomic Structure Generator}\label{sec:cspd}

The Crystal Structure Prototype Database (CSPD)\cite{db_cspd} is constructed by extracting crystal structure prototypes from the Crystallography Open Database (COD)\cite{db_cod}, which includes various compounds except bio-polymers. The process involves filtering structures for quality, classifying them based on composition and atom count, distinguishing between inorganic and organic structures, and assessing structural similarity. 

The CSPD approach, developed by Chuanxun Su \textit{et al.}, generates structures for prediction by first transforming a raw database into a composition-crystal-structure prototype database, significantly reducing the number of candidate structures \cite{db_cspd}. This process, known as the big data method (BDM), involves selecting structure prototypes based on targeted composition, substituting elements, adjusting lattice parameters, and checking minimal interatomic distances. The effectiveness of this method was demonstrated through the generation and evaluation of candidate structures for typical systems, showing that the BDM not only efficiently predicts lower-energy structures but also aligns with experimental findings, underscoring its potential in identifying novel, energetically favorable material configurations.

The CSPD structure generator is publicly available on GitHub \cite{CSPDGithub}. When utilizing the CSPD for structure generation, users can choose to specify either atomic density or volume as guiding parameters for crystal structure creation. When users do not provide these parameters, CSPD employs the covalent radii of species to deduce atomic density, determining the arrangement density of atoms within the generated structure. This approach provides flexibility in tailoring structure generation without requiring explicit pressure settings. In this study, we have used this option to generate crystal samples for the unsynthesizable class. 

Alternative methods to the CSPD have been developed in recent years. One notable examples is the CrystaLLM by Luis Antunes \textit{et al.}, which presents an innovative approach to the generation of crystal structures \cite{antunes2024crystal}. Utilizing autoregressive large language models to interpret Crystallographic Information File formats, CrystaLLM accelerates the discovery of inorganic compounds suitable for applications in energy and electronics, highlighting a significant advancement in the efficiency of crystal structure prediction.

\section{Distribution of ICSD crystals over external pressure}\label{sec:icsd_dist}
\begin{figure*}[h]
    \centering
    \includegraphics[width=0.7\textwidth]{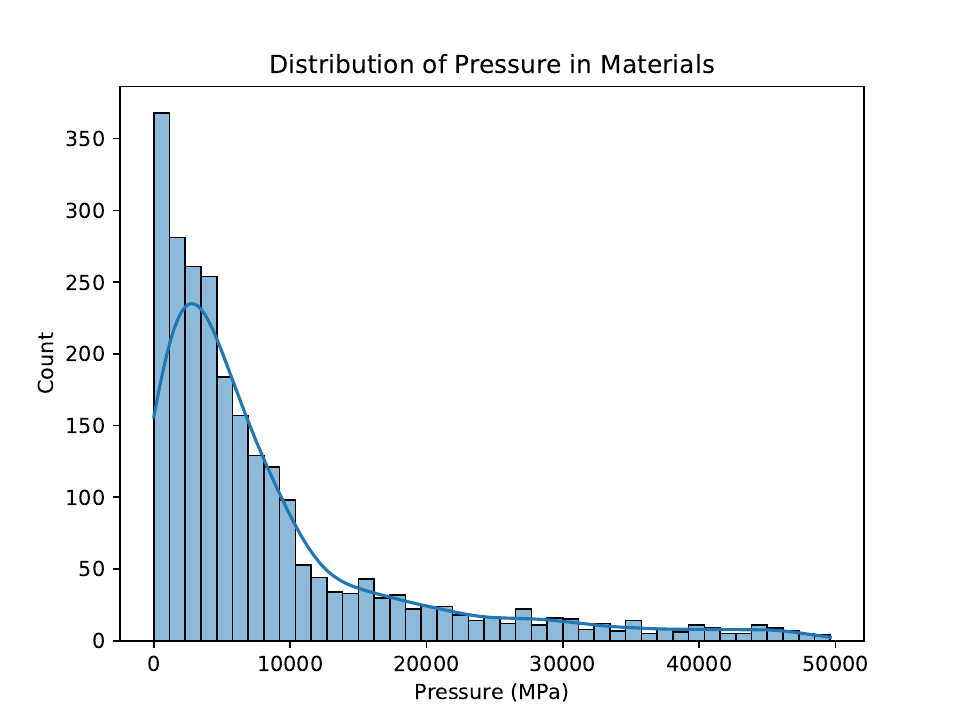}
    \caption[pressure]{
    Distribution for high-pressure ICSD crystal samples (i.e. above ambient pressure). Only samples with pressure below 50 GPa are displayed. The pressure value indicates the external pressure during synthesis.}
    \label{fig:pressure}
\end{figure*}

See Fig. \ref{fig:pressure}. 

\clearpage
\section*{Data availability}
The data developed or used by this study is available on our GitHub repository at \url{https://github.com/kadkhodaei-research-group/XIE-SPP}.

\section*{Code availability}
The codes developed or utilized in this study are openly accessible to support transparency and facilitate further research. They can be found in our GitHub repository at \url{https://github.com/kadkhodaei-research-group/XIE-SPP}.
\section*{Author Contribution Statement}
A.D. and S.K. conceived the research. S.K. and E.Z. supervised the study. A.D. carried out the method development and performed the calculations and analysis. All authors participated in discussing the results and commented on the manuscript.
\section*{Acknowledgements}
This research is based upon work supported by the National Science Foundation (NSF) under Award Numbers DMR-2119308 and DMR-2119065. We used resources at the Electronic Visualization Laboratory (EVL) at University of Illinois Chicago, available through the NSF Award CNS-1828265. 
\section*{Competing Interests} 
All authors declare no financial or non-financial competing interests. 

\clearpage

\begin{thebibliography}{60}%
\makeatletter
\providecommand \@ifxundefined [1]{%
 \@ifx{#1\undefined}
}%
\providecommand \@ifnum [1]{%
 \ifnum #1\expandafter \@firstoftwo
 \else \expandafter \@secondoftwo
 \fi
}%
\providecommand \@ifx [1]{%
 \ifx #1\expandafter \@firstoftwo
 \else \expandafter \@secondoftwo
 \fi
}%
\providecommand \natexlab [1]{#1}%
\providecommand \enquote  [1]{``#1''}%
\providecommand \bibnamefont  [1]{#1}%
\providecommand \bibfnamefont [1]{#1}%
\providecommand \citenamefont [1]{#1}%
\providecommand \href@noop [0]{\@secondoftwo}%
\providecommand \href [0]{\begingroup \@sanitize@url \@href}%
\providecommand \@href[1]{\@@startlink{#1}\@@href}%
\providecommand \@@href[1]{\endgroup#1\@@endlink}%
\providecommand \@sanitize@url [0]{\catcode `\\12\catcode `\$12\catcode
  `\&12\catcode `\#12\catcode `\^12\catcode `\_12\catcode `\%12\relax}%
\providecommand \@@startlink[1]{}%
\providecommand \@@endlink[0]{}%
\providecommand \url  [0]{\begingroup\@sanitize@url \@url }%
\providecommand \@url [1]{\endgroup\@href {#1}{\urlprefix }}%
\providecommand \urlprefix  [0]{URL }%
\providecommand \Eprint [0]{\href }%
\providecommand \doibase [0]{https://doi.org/}%
\providecommand \selectlanguage [0]{\@gobble}%
\providecommand \bibinfo  [0]{\@secondoftwo}%
\providecommand \bibfield  [0]{\@secondoftwo}%
\providecommand \translation [1]{[#1]}%
\providecommand \BibitemOpen [0]{}%
\providecommand \bibitemStop [0]{}%
\providecommand \bibitemNoStop [0]{.\EOS\space}%
\providecommand \EOS [0]{\spacefactor3000\relax}%
\providecommand \BibitemShut  [1]{\csname bibitem#1\endcsname}%
\let\auto@bib@innerbib\@empty
\bibitem [{\citenamefont {Butler}\ \emph {et~al.}(2018)\citenamefont {Butler},
  \citenamefont {Davies}, \citenamefont {Cartwright}, \citenamefont {Isayev},\
  and\ \citenamefont {Walsh}}]{ml_importance_Butler}%
  \BibitemOpen
  \bibfield  {author} {\bibinfo {author} {\bibfnamefont {K.~T.}\ \bibnamefont
  {Butler}}, \bibinfo {author} {\bibfnamefont {D.~W.}\ \bibnamefont {Davies}},
  \bibinfo {author} {\bibfnamefont {H.}~\bibnamefont {Cartwright}}, \bibinfo
  {author} {\bibfnamefont {O.}~\bibnamefont {Isayev}},\ and\ \bibinfo {author}
  {\bibfnamefont {A.}~\bibnamefont {Walsh}},\ }\href
  {https://doi.org/10.1038/s41586-018-0337-2} {\bibfield  {journal} {\bibinfo
  {journal} {Nature}\ }\textbf {\bibinfo {volume} {559}},\ \bibinfo {pages}
  {547} (\bibinfo {year} {2018})}\BibitemShut {NoStop}%
\bibitem [{\citenamefont {Himanen}\ \emph {et~al.}(2019)\citenamefont
  {Himanen}, \citenamefont {Geurts}, \citenamefont {Foster},\ and\
  \citenamefont {Rinke}}]{HimanenGeurtsFosterEtAl2019}%
  \BibitemOpen
  \bibfield  {author} {\bibinfo {author} {\bibfnamefont {L.}~\bibnamefont
  {Himanen}}, \bibinfo {author} {\bibfnamefont {A.}~\bibnamefont {Geurts}},
  \bibinfo {author} {\bibfnamefont {A.~S.}\ \bibnamefont {Foster}},\ and\
  \bibinfo {author} {\bibfnamefont {P.}~\bibnamefont {Rinke}},\ }\href
  {https://doi.org/10.1002/advs.201900808} {\bibfield  {journal} {\bibinfo
  {journal} {Advanced Science}\ }\textbf {\bibinfo {volume} {articlenumber
  1900808}},\ \bibinfo {pages} {23} (\bibinfo {year} {2019})}\BibitemShut
  {NoStop}%
\bibitem [{\citenamefont {de~Pablo}\ \emph {et~al.}(2019)\citenamefont
  {de~Pablo}, \citenamefont {Jackson}, \citenamefont {Webb}, \citenamefont
  {Chen}, \citenamefont {Moore}, \citenamefont {Morgan}, \citenamefont
  {Jacobs}, \citenamefont {Pollock}, \citenamefont {Schlom}, \citenamefont
  {Toberer}, \citenamefont {Analytis}, \citenamefont {Dabo}, \citenamefont
  {DeLongchamp}, \citenamefont {Fiete}, \citenamefont {Grason}, \citenamefont
  {Hautier}, \citenamefont {Mo}, \citenamefont {Rajan}, \citenamefont {Reed},
  \citenamefont {Rodriguez}, \citenamefont {Stevanovic}, \citenamefont
  {Suntivich}, \citenamefont {Thornton},\ and\ \citenamefont
  {Zhao}}]{Pablo:2019aa}%
  \BibitemOpen
  \bibfield  {author} {\bibinfo {author} {\bibfnamefont {J.~J.}\ \bibnamefont
  {de~Pablo}}, \bibinfo {author} {\bibfnamefont {N.~E.}\ \bibnamefont
  {Jackson}}, \bibinfo {author} {\bibfnamefont {M.~A.}\ \bibnamefont {Webb}},
  \bibinfo {author} {\bibfnamefont {L.-Q.}\ \bibnamefont {Chen}}, \bibinfo
  {author} {\bibfnamefont {J.~E.}\ \bibnamefont {Moore}}, \bibinfo {author}
  {\bibfnamefont {D.}~\bibnamefont {Morgan}}, \bibinfo {author} {\bibfnamefont
  {R.}~\bibnamefont {Jacobs}}, \bibinfo {author} {\bibfnamefont
  {T.}~\bibnamefont {Pollock}}, \bibinfo {author} {\bibfnamefont {D.~G.}\
  \bibnamefont {Schlom}}, \bibinfo {author} {\bibfnamefont {E.~S.}\
  \bibnamefont {Toberer}}, \bibinfo {author} {\bibfnamefont {J.}~\bibnamefont
  {Analytis}}, \bibinfo {author} {\bibfnamefont {I.}~\bibnamefont {Dabo}},
  \bibinfo {author} {\bibfnamefont {D.~M.}\ \bibnamefont {DeLongchamp}},
  \bibinfo {author} {\bibfnamefont {G.~A.}\ \bibnamefont {Fiete}}, \bibinfo
  {author} {\bibfnamefont {G.~M.}\ \bibnamefont {Grason}}, \bibinfo {author}
  {\bibfnamefont {G.}~\bibnamefont {Hautier}}, \bibinfo {author} {\bibfnamefont
  {Y.}~\bibnamefont {Mo}}, \bibinfo {author} {\bibfnamefont {K.}~\bibnamefont
  {Rajan}}, \bibinfo {author} {\bibfnamefont {E.~J.}\ \bibnamefont {Reed}},
  \bibinfo {author} {\bibfnamefont {E.}~\bibnamefont {Rodriguez}}, \bibinfo
  {author} {\bibfnamefont {V.}~\bibnamefont {Stevanovic}}, \bibinfo {author}
  {\bibfnamefont {J.}~\bibnamefont {Suntivich}}, \bibinfo {author}
  {\bibfnamefont {K.}~\bibnamefont {Thornton}},\ and\ \bibinfo {author}
  {\bibfnamefont {J.-C.}\ \bibnamefont {Zhao}},\ }\href
  {https://doi.org/10.1038/s41524-019-0173-4} {\bibfield  {journal} {\bibinfo
  {journal} {npj Computational Materials}\ }\textbf {\bibinfo {volume} {5}},\
  \bibinfo {pages} {41} (\bibinfo {year} {2019})}\BibitemShut {NoStop}%
\bibitem [{\citenamefont {Tian}\ \emph {et~al.}(2021)\citenamefont {Tian},
  \citenamefont {Xue}, \citenamefont {Yuan}, \citenamefont {Zhou},
  \citenamefont {Ding}, \citenamefont {Sun},\ and\ \citenamefont
  {Lookman}}]{Tian2021}%
  \BibitemOpen
  \bibfield  {author} {\bibinfo {author} {\bibfnamefont {Y.}~\bibnamefont
  {Tian}}, \bibinfo {author} {\bibfnamefont {D.}~\bibnamefont {Xue}}, \bibinfo
  {author} {\bibfnamefont {R.}~\bibnamefont {Yuan}}, \bibinfo {author}
  {\bibfnamefont {Y.}~\bibnamefont {Zhou}}, \bibinfo {author} {\bibfnamefont
  {X.}~\bibnamefont {Ding}}, \bibinfo {author} {\bibfnamefont {J.}~\bibnamefont
  {Sun}},\ and\ \bibinfo {author} {\bibfnamefont {T.}~\bibnamefont {Lookman}},\
  }\href {https://doi.org/10.1103/PhysRevMaterials.5.013802} {\bibfield
  {journal} {\bibinfo  {journal} {Phys. Rev. Mater.}\ }\textbf {\bibinfo
  {volume} {5}},\ \bibinfo {pages} {013802} (\bibinfo {year}
  {2021})}\BibitemShut {NoStop}%
\bibitem [{\citenamefont {Oganov}\ \emph {et~al.}(2019)\citenamefont {Oganov},
  \citenamefont {Pickard}, \citenamefont {Zhu},\ and\ \citenamefont
  {Needs}}]{Oganov2019}%
  \BibitemOpen
  \bibfield  {author} {\bibinfo {author} {\bibfnamefont {A.~R.}\ \bibnamefont
  {Oganov}}, \bibinfo {author} {\bibfnamefont {C.~J.}\ \bibnamefont {Pickard}},
  \bibinfo {author} {\bibfnamefont {Q.}~\bibnamefont {Zhu}},\ and\ \bibinfo
  {author} {\bibfnamefont {R.~J.}\ \bibnamefont {Needs}},\ }\href
  {https://doi.org/10.1038/s41578-019-0101-8} {\bibfield  {journal} {\bibinfo
  {journal} {Nature Reviews Materials}\ }\textbf {\bibinfo {volume} {4}},\
  \bibinfo {pages} {331} (\bibinfo {year} {2019})}\BibitemShut {NoStop}%
\bibitem [{\citenamefont {Umehara}\ \emph {et~al.}(2019)\citenamefont
  {Umehara}, \citenamefont {Stein}, \citenamefont {Guevarra}, \citenamefont
  {Newhouse}, \citenamefont {Boyd},\ and\ \citenamefont
  {Gregoire}}]{Umehara2019}%
  \BibitemOpen
  \bibfield  {author} {\bibinfo {author} {\bibfnamefont {M.}~\bibnamefont
  {Umehara}}, \bibinfo {author} {\bibfnamefont {H.~S.}\ \bibnamefont {Stein}},
  \bibinfo {author} {\bibfnamefont {D.}~\bibnamefont {Guevarra}}, \bibinfo
  {author} {\bibfnamefont {P.~F.}\ \bibnamefont {Newhouse}}, \bibinfo {author}
  {\bibfnamefont {D.~A.}\ \bibnamefont {Boyd}},\ and\ \bibinfo {author}
  {\bibfnamefont {J.~M.}\ \bibnamefont {Gregoire}},\ }\href
  {https://doi.org/10.1038/s41524-019-0172-5} {\bibfield  {journal} {\bibinfo
  {journal} {npj Computational Materials}\ }\textbf {\bibinfo {volume} {5}},\
  \bibinfo {pages} {34} (\bibinfo {year} {2019})}\BibitemShut {NoStop}%
\bibitem [{\citenamefont {Jablonka}\ \emph {et~al.}(2020)\citenamefont
  {Jablonka}, \citenamefont {Ongari}, \citenamefont {Moosavi},\ and\
  \citenamefont {Smit}}]{Jablonka2020}%
  \BibitemOpen
  \bibfield  {author} {\bibinfo {author} {\bibfnamefont {K.~M.}\ \bibnamefont
  {Jablonka}}, \bibinfo {author} {\bibfnamefont {D.}~\bibnamefont {Ongari}},
  \bibinfo {author} {\bibfnamefont {S.~M.}\ \bibnamefont {Moosavi}},\ and\
  \bibinfo {author} {\bibfnamefont {B.}~\bibnamefont {Smit}},\ }\href
  {https://doi.org/10.1021/acs.chemrev.0c00004} {\bibfield  {journal} {\bibinfo
   {journal} {Chemical Reviews}\ }\textbf {\bibinfo {volume} {120}},\ \bibinfo
  {pages} {8066} (\bibinfo {year} {2020})}\BibitemShut {NoStop}%
\bibitem [{\citenamefont {Himanen}\ \emph {et~al.}(2020)\citenamefont
  {Himanen}, \citenamefont {Jäger}, \citenamefont {Morooka}, \citenamefont
  {{Federici Canova}}, \citenamefont {Ranawat}, \citenamefont {Gao},
  \citenamefont {Rinke},\ and\ \citenamefont {Foster}}]{HIMANEN2020106949}%
  \BibitemOpen
  \bibfield  {author} {\bibinfo {author} {\bibfnamefont {L.}~\bibnamefont
  {Himanen}}, \bibinfo {author} {\bibfnamefont {M.~O.}\ \bibnamefont {Jäger}},
  \bibinfo {author} {\bibfnamefont {E.~V.}\ \bibnamefont {Morooka}}, \bibinfo
  {author} {\bibfnamefont {F.}~\bibnamefont {{Federici Canova}}}, \bibinfo
  {author} {\bibfnamefont {Y.~S.}\ \bibnamefont {Ranawat}}, \bibinfo {author}
  {\bibfnamefont {D.~Z.}\ \bibnamefont {Gao}}, \bibinfo {author} {\bibfnamefont
  {P.}~\bibnamefont {Rinke}},\ and\ \bibinfo {author} {\bibfnamefont {A.~S.}\
  \bibnamefont {Foster}},\ }\href
  {https://doi.org/https://doi.org/10.1016/j.cpc.2019.106949} {\bibfield
  {journal} {\bibinfo  {journal} {Computer Physics Communications}\ }\textbf
  {\bibinfo {volume} {247}},\ \bibinfo {pages} {106949} (\bibinfo {year}
  {2020})}\BibitemShut {NoStop}%
\bibitem [{\citenamefont {Morgan}\ and\ \citenamefont
  {Jacobs}(2020)}]{Morgan2020}%
  \BibitemOpen
  \bibfield  {author} {\bibinfo {author} {\bibfnamefont {D.}~\bibnamefont
  {Morgan}}\ and\ \bibinfo {author} {\bibfnamefont {R.}~\bibnamefont
  {Jacobs}},\ }\href {https://doi.org/10.1146/annurev-matsci-070218-010015}
  {\bibfield  {journal} {\bibinfo  {journal} {Annual Review of Materials
  Research}\ }\textbf {\bibinfo {volume} {50}},\ \bibinfo {pages} {71}
  (\bibinfo {year} {2020})},\ \Eprint
  {https://arxiv.org/abs/https://doi.org/10.1146/annurev-matsci-070218-010015}
  {https://doi.org/10.1146/annurev-matsci-070218-010015} \BibitemShut {NoStop}%
\bibitem [{\citenamefont {Hart}\ \emph {et~al.}(2021)\citenamefont {Hart},
  \citenamefont {Mueller}, \citenamefont {Toher},\ and\ \citenamefont
  {Curtarolo}}]{Hart2021}%
  \BibitemOpen
  \bibfield  {author} {\bibinfo {author} {\bibfnamefont {G.~L.~W.}\
  \bibnamefont {Hart}}, \bibinfo {author} {\bibfnamefont {T.}~\bibnamefont
  {Mueller}}, \bibinfo {author} {\bibfnamefont {C.}~\bibnamefont {Toher}},\
  and\ \bibinfo {author} {\bibfnamefont {S.}~\bibnamefont {Curtarolo}},\ }\href
  {https://doi.org/10.1038/s41578-021-00340-w} {\bibfield  {journal} {\bibinfo
  {journal} {Nature Reviews Materials}\ }\textbf {\bibinfo {volume} {6}},\
  \bibinfo {pages} {730} (\bibinfo {year} {2021})}\BibitemShut {NoStop}%
\bibitem [{\citenamefont {Gong}\ \emph {et~al.}(2021)\citenamefont {Gong},
  \citenamefont {Wang}, \citenamefont {Zhu}, \citenamefont {Chen},
  \citenamefont {Yang}, \citenamefont {Buehler}, \citenamefont {Shao-Horn},\
  and\ \citenamefont {Grossman}}]{Gong2021}%
  \BibitemOpen
  \bibfield  {author} {\bibinfo {author} {\bibfnamefont {S.}~\bibnamefont
  {Gong}}, \bibinfo {author} {\bibfnamefont {S.}~\bibnamefont {Wang}}, \bibinfo
  {author} {\bibfnamefont {T.}~\bibnamefont {Zhu}}, \bibinfo {author}
  {\bibfnamefont {X.}~\bibnamefont {Chen}}, \bibinfo {author} {\bibfnamefont
  {Z.}~\bibnamefont {Yang}}, \bibinfo {author} {\bibfnamefont {M.~J.}\
  \bibnamefont {Buehler}}, \bibinfo {author} {\bibfnamefont {Y.}~\bibnamefont
  {Shao-Horn}},\ and\ \bibinfo {author} {\bibfnamefont {J.~C.}\ \bibnamefont
  {Grossman}},\ }\href {https://doi.org/10.1021/jacsau.1c00260} {\bibfield
  {journal} {\bibinfo  {journal} {JACS Au}\ }\textbf {\bibinfo {volume} {1}},\
  \bibinfo {pages} {1904} (\bibinfo {year} {2021})},\ \Eprint
  {https://arxiv.org/abs/https://doi.org/10.1021/jacsau.1c00260}
  {https://doi.org/10.1021/jacsau.1c00260} \BibitemShut {NoStop}%
\bibitem [{\citenamefont {Zhong}\ \emph {et~al.}(2022)\citenamefont {Zhong},
  \citenamefont {Gallagher}, \citenamefont {Liu}, \citenamefont {Kailkhura},
  \citenamefont {Hiszpanski},\ and\ \citenamefont {Han}}]{Zhong2022}%
  \BibitemOpen
  \bibfield  {author} {\bibinfo {author} {\bibfnamefont {X.}~\bibnamefont
  {Zhong}}, \bibinfo {author} {\bibfnamefont {B.}~\bibnamefont {Gallagher}},
  \bibinfo {author} {\bibfnamefont {S.}~\bibnamefont {Liu}}, \bibinfo {author}
  {\bibfnamefont {B.}~\bibnamefont {Kailkhura}}, \bibinfo {author}
  {\bibfnamefont {A.}~\bibnamefont {Hiszpanski}},\ and\ \bibinfo {author}
  {\bibfnamefont {T.~Y.-J.}\ \bibnamefont {Han}},\ }\href
  {https://doi.org/10.1038/s41524-022-00884-7} {\bibfield  {journal} {\bibinfo
  {journal} {npj Computational Materials}\ }\textbf {\bibinfo {volume} {8}},\
  \bibinfo {pages} {204} (\bibinfo {year} {2022})}\BibitemShut {NoStop}%
\bibitem [{\citenamefont {Damewood}\ \emph {et~al.}(2023)\citenamefont
  {Damewood}, \citenamefont {Karaguesian}, \citenamefont {Lunger},
  \citenamefont {Tan}, \citenamefont {Xie}, \citenamefont {Peng},\ and\
  \citenamefont {G\'{o}mez-Bombarelli}}]{Damewood2023}%
  \BibitemOpen
  \bibfield  {author} {\bibinfo {author} {\bibfnamefont {J.}~\bibnamefont
  {Damewood}}, \bibinfo {author} {\bibfnamefont {J.}~\bibnamefont
  {Karaguesian}}, \bibinfo {author} {\bibfnamefont {J.~R.}\ \bibnamefont
  {Lunger}}, \bibinfo {author} {\bibfnamefont {A.~R.}\ \bibnamefont {Tan}},
  \bibinfo {author} {\bibfnamefont {M.}~\bibnamefont {Xie}}, \bibinfo {author}
  {\bibfnamefont {J.}~\bibnamefont {Peng}},\ and\ \bibinfo {author}
  {\bibfnamefont {R.}~\bibnamefont {G\'{o}mez-Bombarelli}},\ }\href
  {https://doi.org/10.1146/annurev-matsci-080921-085947} {\bibfield  {journal}
  {\bibinfo  {journal} {Annual Review of Materials Research}\ }\textbf
  {\bibinfo {volume} {53}},\ \bibinfo {pages} {399} (\bibinfo {year} {2023})},\
  \Eprint
  {https://arxiv.org/abs/https://doi.org/10.1146/annurev-matsci-080921-085947}
  {https://doi.org/10.1146/annurev-matsci-080921-085947} \BibitemShut {NoStop}%
\bibitem [{\citenamefont {Xu}\ \emph {et~al.}(2023)\citenamefont {Xu},
  \citenamefont {Ji}, \citenamefont {Li},\ and\ \citenamefont {Lu}}]{Xu2023}%
  \BibitemOpen
  \bibfield  {author} {\bibinfo {author} {\bibfnamefont {P.}~\bibnamefont
  {Xu}}, \bibinfo {author} {\bibfnamefont {X.}~\bibnamefont {Ji}}, \bibinfo
  {author} {\bibfnamefont {M.}~\bibnamefont {Li}},\ and\ \bibinfo {author}
  {\bibfnamefont {W.}~\bibnamefont {Lu}},\ }\href
  {https://doi.org/10.1038/s41524-023-01000-z} {\bibfield  {journal} {\bibinfo
  {journal} {npj Computational Materials}\ }\textbf {\bibinfo {volume} {9}},\
  \bibinfo {pages} {42} (\bibinfo {year} {2023})}\BibitemShut {NoStop}%
\bibitem [{\citenamefont {Agrawal}\ and\ \citenamefont
  {Choudhary}(2016)}]{Agrawal2016}%
  \BibitemOpen
  \bibfield  {author} {\bibinfo {author} {\bibfnamefont {A.}~\bibnamefont
  {Agrawal}}\ and\ \bibinfo {author} {\bibfnamefont {A.}~\bibnamefont
  {Choudhary}},\ }\bibfield  {journal} {\bibinfo  {journal} {APL Materials}\
  }\textbf {\bibinfo {volume} {4}},\ \href {https://doi.org/10.1063/1.4946894}
  {10.1063/1.4946894} (\bibinfo {year} {2016})\BibitemShut {NoStop}%
\bibitem [{\citenamefont {Liang}\ \emph {et~al.}(2022)\citenamefont {Liang},
  \citenamefont {Tadesse}, \citenamefont {Ho}, \citenamefont {Fei-Fei},
  \citenamefont {Zaharia}, \citenamefont {Zhang},\ and\ \citenamefont
  {Zou}}]{databias_liang2022}%
  \BibitemOpen
  \bibfield  {author} {\bibinfo {author} {\bibfnamefont {W.}~\bibnamefont
  {Liang}}, \bibinfo {author} {\bibfnamefont {G.~A.}\ \bibnamefont {Tadesse}},
  \bibinfo {author} {\bibfnamefont {D.}~\bibnamefont {Ho}}, \bibinfo {author}
  {\bibfnamefont {L.}~\bibnamefont {Fei-Fei}}, \bibinfo {author} {\bibfnamefont
  {M.}~\bibnamefont {Zaharia}}, \bibinfo {author} {\bibfnamefont
  {C.}~\bibnamefont {Zhang}},\ and\ \bibinfo {author} {\bibfnamefont
  {J.}~\bibnamefont {Zou}},\ }\href
  {https://doi.org/10.1038/s42256-022-00516-1} {\bibfield  {journal} {\bibinfo
  {journal} {Nature Machine Intelligence}\ }\textbf {\bibinfo {volume} {4}},\
  \bibinfo {pages} {669} (\bibinfo {year} {2022})}\BibitemShut {NoStop}%
\bibitem [{\citenamefont {Davariashtiyani}\ \emph {et~al.}(2021)\citenamefont
  {Davariashtiyani}, \citenamefont {Kadkhodaie},\ and\ \citenamefont
  {Kadkhodaei}}]{Davariashtiyani_2021_SYN}%
  \BibitemOpen
  \bibfield  {author} {\bibinfo {author} {\bibfnamefont {A.}~\bibnamefont
  {Davariashtiyani}}, \bibinfo {author} {\bibfnamefont {Z.}~\bibnamefont
  {Kadkhodaie}},\ and\ \bibinfo {author} {\bibfnamefont {S.}~\bibnamefont
  {Kadkhodaei}},\ }\href {https://doi.org/10.1038/s43246-021-00219-x}
  {\bibfield  {journal} {\bibinfo  {journal} {Communications Materials}\
  }\textbf {\bibinfo {volume} {2}},\ \bibinfo {pages} {115} (\bibinfo {year}
  {2021})}\BibitemShut {NoStop}%
\bibitem [{\citenamefont {Frey}\ \emph {et~al.}(2019)\citenamefont {Frey},
  \citenamefont {Wang}, \citenamefont {Vega~Bellido}, \citenamefont {Anasori},
  \citenamefont {Gogotsi},\ and\ \citenamefont {Shenoy}}]{syn_Frey2019}%
  \BibitemOpen
  \bibfield  {author} {\bibinfo {author} {\bibfnamefont {N.~C.}\ \bibnamefont
  {Frey}}, \bibinfo {author} {\bibfnamefont {J.}~\bibnamefont {Wang}}, \bibinfo
  {author} {\bibfnamefont {G.~I.}\ \bibnamefont {Vega~Bellido}}, \bibinfo
  {author} {\bibfnamefont {B.}~\bibnamefont {Anasori}}, \bibinfo {author}
  {\bibfnamefont {Y.}~\bibnamefont {Gogotsi}},\ and\ \bibinfo {author}
  {\bibfnamefont {V.~B.}\ \bibnamefont {Shenoy}},\ }\href
  {https://doi.org/10.1021/acsnano.8b08014} {\bibfield  {journal} {\bibinfo
  {journal} {ACS Nano}\ }\textbf {\bibinfo {volume} {13}},\ \bibinfo {pages}
  {3031} (\bibinfo {year} {2019})}\BibitemShut {NoStop}%
\bibitem [{\citenamefont {Jang}\ \emph {et~al.}(2020)\citenamefont {Jang},
  \citenamefont {Gu}, \citenamefont {Noh}, \citenamefont {Kim},\ and\
  \citenamefont {Jung}}]{syn_Jang2020}%
  \BibitemOpen
  \bibfield  {author} {\bibinfo {author} {\bibfnamefont {J.}~\bibnamefont
  {Jang}}, \bibinfo {author} {\bibfnamefont {G.~H.}\ \bibnamefont {Gu}},
  \bibinfo {author} {\bibfnamefont {J.}~\bibnamefont {Noh}}, \bibinfo {author}
  {\bibfnamefont {J.}~\bibnamefont {Kim}},\ and\ \bibinfo {author}
  {\bibfnamefont {Y.}~\bibnamefont {Jung}},\ }\href
  {https://doi.org/10.1021/jacs.0c07384} {\bibfield  {journal} {\bibinfo
  {journal} {Journal of the American Chemical Society}\ }\textbf {\bibinfo
  {volume} {142}},\ \bibinfo {pages} {18836} (\bibinfo {year}
  {2020})}\BibitemShut {NoStop}%
\bibitem [{\citenamefont {Antoniuk}\ \emph {et~al.}(2023)\citenamefont
  {Antoniuk}, \citenamefont {Cheon}, \citenamefont {Wang}, \citenamefont
  {Bernstein}, \citenamefont {Cai},\ and\ \citenamefont
  {Reed}}]{syn_Antoniuk2023}%
  \BibitemOpen
  \bibfield  {author} {\bibinfo {author} {\bibfnamefont {E.~R.}\ \bibnamefont
  {Antoniuk}}, \bibinfo {author} {\bibfnamefont {G.}~\bibnamefont {Cheon}},
  \bibinfo {author} {\bibfnamefont {G.}~\bibnamefont {Wang}}, \bibinfo {author}
  {\bibfnamefont {D.}~\bibnamefont {Bernstein}}, \bibinfo {author}
  {\bibfnamefont {W.}~\bibnamefont {Cai}},\ and\ \bibinfo {author}
  {\bibfnamefont {E.~J.}\ \bibnamefont {Reed}},\ }\href
  {https://doi.org/10.1038/s41524-023-01114-4} {\bibfield  {journal} {\bibinfo
  {journal} {npj Computational Materials}\ }\textbf {\bibinfo {volume} {9}},\
  \bibinfo {pages} {155} (\bibinfo {year} {2023})}\BibitemShut {NoStop}%
\bibitem [{\citenamefont {Gleaves}\ \emph {et~al.}(2023)\citenamefont
  {Gleaves}, \citenamefont {Fu}, \citenamefont {Dilanga~Siriwardane},
  \citenamefont {Zhao},\ and\ \citenamefont {Hu}}]{syn_Gleaves2023}%
  \BibitemOpen
  \bibfield  {author} {\bibinfo {author} {\bibfnamefont {D.}~\bibnamefont
  {Gleaves}}, \bibinfo {author} {\bibfnamefont {N.}~\bibnamefont {Fu}},
  \bibinfo {author} {\bibfnamefont {E.~M.}\ \bibnamefont
  {Dilanga~Siriwardane}}, \bibinfo {author} {\bibfnamefont {Y.}~\bibnamefont
  {Zhao}},\ and\ \bibinfo {author} {\bibfnamefont {J.}~\bibnamefont {Hu}},\
  }\href {https://doi.org/10.1039/D2DD00098A} {\bibfield  {journal} {\bibinfo
  {journal} {Digital Discovery}\ }\textbf {\bibinfo {volume} {2}},\ \bibinfo
  {pages} {377} (\bibinfo {year} {2023})}\BibitemShut {NoStop}%
\bibitem [{\citenamefont {Gong}\ \emph {et~al.}(2022)\citenamefont {Gong},
  \citenamefont {Wang}, \citenamefont {Liu}, \citenamefont {Han}, \citenamefont
  {You}, \citenamefont {Yang},\ and\ \citenamefont {Tao}}]{ml_PU}%
  \BibitemOpen
  \bibfield  {author} {\bibinfo {author} {\bibfnamefont {C.}~\bibnamefont
  {Gong}}, \bibinfo {author} {\bibfnamefont {Q.}~\bibnamefont {Wang}}, \bibinfo
  {author} {\bibfnamefont {T.}~\bibnamefont {Liu}}, \bibinfo {author}
  {\bibfnamefont {B.}~\bibnamefont {Han}}, \bibinfo {author} {\bibfnamefont
  {J.}~\bibnamefont {You}}, \bibinfo {author} {\bibfnamefont {J.}~\bibnamefont
  {Yang}},\ and\ \bibinfo {author} {\bibfnamefont {D.}~\bibnamefont {Tao}},\
  }\href {https://doi.org/10.1109/TPAMI.2021.3061456} {\bibfield  {journal}
  {\bibinfo  {journal} {IEEE Transactions on Pattern Analysis and Machine
  Intelligence}\ }\textbf {\bibinfo {volume} {44}},\ \bibinfo {pages} {4163}
  (\bibinfo {year} {2022})}\BibitemShut {NoStop}%
\bibitem [{\citenamefont {Kumagai}\ \emph {et~al.}(2022)\citenamefont
  {Kumagai}, \citenamefont {Ando}, \citenamefont {Tanaka}, \citenamefont
  {Tsuda}, \citenamefont {Katsura},\ and\ \citenamefont
  {Kurosaki}}]{data_bias_Kumagai2022}%
  \BibitemOpen
  \bibfield  {author} {\bibinfo {author} {\bibfnamefont {M.}~\bibnamefont
  {Kumagai}}, \bibinfo {author} {\bibfnamefont {Y.}~\bibnamefont {Ando}},
  \bibinfo {author} {\bibfnamefont {A.}~\bibnamefont {Tanaka}}, \bibinfo
  {author} {\bibfnamefont {K.}~\bibnamefont {Tsuda}}, \bibinfo {author}
  {\bibfnamefont {Y.}~\bibnamefont {Katsura}},\ and\ \bibinfo {author}
  {\bibfnamefont {K.}~\bibnamefont {Kurosaki}},\ }\href
  {https://doi.org/10.1080/27660400.2022.2109447} {\bibfield  {journal}
  {\bibinfo  {journal} {Science and Technology of Advanced Materials: Methods}\
  }\textbf {\bibinfo {volume} {2}},\ \bibinfo {pages} {302} (\bibinfo {year}
  {2022})},\ \Eprint
  {https://arxiv.org/abs/https://doi.org/10.1080/27660400.2022.2109447}
  {https://doi.org/10.1080/27660400.2022.2109447} \BibitemShut {NoStop}%
\bibitem [{\citenamefont {Zhang}\ \emph {et~al.}(2023)\citenamefont {Zhang},
  \citenamefont {Chen}, \citenamefont {Rondinelli},\ and\ \citenamefont
  {Chen}}]{data_bias_zhang2023entropy}%
  \BibitemOpen
  \bibfield  {author} {\bibinfo {author} {\bibfnamefont {H.}~\bibnamefont
  {Zhang}}, \bibinfo {author} {\bibfnamefont {W.~W.}\ \bibnamefont {Chen}},
  \bibinfo {author} {\bibfnamefont {J.~M.}\ \bibnamefont {Rondinelli}},\ and\
  \bibinfo {author} {\bibfnamefont {W.}~\bibnamefont {Chen}},\ }\bibfield
  {journal} {\bibinfo  {journal} {Applied Physics Reviews}\ }\textbf {\bibinfo
  {volume} {10}},\ \href {https://doi.org/10.1063/5.0138913}
  {10.1063/5.0138913} (\bibinfo {year} {2023})\BibitemShut {NoStop}%
\bibitem [{\citenamefont {Li}\ \emph {et~al.}(2023)\citenamefont {Li},
  \citenamefont {DeCost}, \citenamefont {Choudhary}, \citenamefont
  {Greenwood},\ and\ \citenamefont {Hattrick-Simpers}}]{data_bias_Li2023}%
  \BibitemOpen
  \bibfield  {author} {\bibinfo {author} {\bibfnamefont {K.}~\bibnamefont
  {Li}}, \bibinfo {author} {\bibfnamefont {B.}~\bibnamefont {DeCost}}, \bibinfo
  {author} {\bibfnamefont {K.}~\bibnamefont {Choudhary}}, \bibinfo {author}
  {\bibfnamefont {M.}~\bibnamefont {Greenwood}},\ and\ \bibinfo {author}
  {\bibfnamefont {J.}~\bibnamefont {Hattrick-Simpers}},\ }\href
  {https://doi.org/10.1038/s41524-023-01012-9} {\bibfield  {journal} {\bibinfo
  {journal} {npj Computational Materials}\ }\textbf {\bibinfo {volume} {9}},\
  \bibinfo {pages} {55} (\bibinfo {year} {2023})}\BibitemShut {NoStop}%
\bibitem [{\citenamefont {Zhang}\ and\ \citenamefont {Ling}(2018)}]{Zhang2018}%
  \BibitemOpen
  \bibfield  {author} {\bibinfo {author} {\bibfnamefont {Y.}~\bibnamefont
  {Zhang}}\ and\ \bibinfo {author} {\bibfnamefont {C.}~\bibnamefont {Ling}},\
  }\href {https://doi.org/10.1038/s41524-018-0081-z} {\bibfield  {journal}
  {\bibinfo  {journal} {npj Computational Materials}\ }\textbf {\bibinfo
  {volume} {4}},\ \bibinfo {pages} {25} (\bibinfo {year} {2018})}\BibitemShut
  {NoStop}%
\bibitem [{\citenamefont {Breuck}\ \emph {et~al.}(2021)\citenamefont {Breuck},
  \citenamefont {Evans},\ and\ \citenamefont {Rignanese}}]{De_Breuck_2021}%
  \BibitemOpen
  \bibfield  {author} {\bibinfo {author} {\bibfnamefont {P.-P.~D.}\
  \bibnamefont {Breuck}}, \bibinfo {author} {\bibfnamefont {M.~L.}\
  \bibnamefont {Evans}},\ and\ \bibinfo {author} {\bibfnamefont {G.-M.}\
  \bibnamefont {Rignanese}},\ }\href {https://doi.org/10.1088/1361-648X/ac1280}
  {\bibfield  {journal} {\bibinfo  {journal} {Journal of Physics: Condensed
  Matter}\ }\textbf {\bibinfo {volume} {33}},\ \bibinfo {pages} {404002}
  (\bibinfo {year} {2021})}\BibitemShut {NoStop}%
\bibitem [{\citenamefont {Saal}\ \emph {et~al.}(2013)\citenamefont {Saal},
  \citenamefont {Kirklin}, \citenamefont {Aykol}, \citenamefont {Meredig},\
  and\ \citenamefont {Wolverton}}]{db_oqmd_1}%
  \BibitemOpen
  \bibfield  {author} {\bibinfo {author} {\bibfnamefont {J.~E.}\ \bibnamefont
  {Saal}}, \bibinfo {author} {\bibfnamefont {S.}~\bibnamefont {Kirklin}},
  \bibinfo {author} {\bibfnamefont {M.}~\bibnamefont {Aykol}}, \bibinfo
  {author} {\bibfnamefont {B.}~\bibnamefont {Meredig}},\ and\ \bibinfo {author}
  {\bibfnamefont {C.}~\bibnamefont {Wolverton}},\ }\href
  {https://doi.org/10.1007/s11837-013-0755-4} {\bibfield  {journal} {\bibinfo
  {journal} {JOM}\ }\textbf {\bibinfo {volume} {65}},\ \bibinfo {pages} {1501}
  (\bibinfo {year} {2013})}\BibitemShut {NoStop}%
\bibitem [{\citenamefont {Kirklin}\ \emph {et~al.}(2015)\citenamefont
  {Kirklin}, \citenamefont {Saal}, \citenamefont {Meredig}, \citenamefont
  {Thompson}, \citenamefont {Doak}, \citenamefont {Aykol}, \citenamefont
  {R{\"u}hl},\ and\ \citenamefont {Wolverton}}]{db_oqmd_2}%
  \BibitemOpen
  \bibfield  {author} {\bibinfo {author} {\bibfnamefont {S.}~\bibnamefont
  {Kirklin}}, \bibinfo {author} {\bibfnamefont {J.~E.}\ \bibnamefont {Saal}},
  \bibinfo {author} {\bibfnamefont {B.}~\bibnamefont {Meredig}}, \bibinfo
  {author} {\bibfnamefont {A.}~\bibnamefont {Thompson}}, \bibinfo {author}
  {\bibfnamefont {J.~W.}\ \bibnamefont {Doak}}, \bibinfo {author}
  {\bibfnamefont {M.}~\bibnamefont {Aykol}}, \bibinfo {author} {\bibfnamefont
  {S.}~\bibnamefont {R{\"u}hl}},\ and\ \bibinfo {author} {\bibfnamefont
  {C.}~\bibnamefont {Wolverton}},\ }\href
  {https://doi.org/10.1038/npjcompumats.2015.10} {\bibfield  {journal}
  {\bibinfo  {journal} {npj Computational Materials}\ }\textbf {\bibinfo
  {volume} {1}},\ \bibinfo {pages} {15010} (\bibinfo {year}
  {2015})}\BibitemShut {NoStop}%
\bibitem [{\citenamefont {Choudhary}\ \emph {et~al.}(2020)\citenamefont
  {Choudhary}, \citenamefont {Garrity}, \citenamefont {Reid}, \citenamefont
  {DeCost}, \citenamefont {Biacchi}, \citenamefont {Hight~Walker},
  \citenamefont {Trautt}, \citenamefont {Hattrick-Simpers}, \citenamefont
  {Kusne}, \citenamefont {Centrone}, \citenamefont {Davydov}, \citenamefont
  {Jiang}, \citenamefont {Pachter}, \citenamefont {Cheon}, \citenamefont
  {Reed}, \citenamefont {Agrawal}, \citenamefont {Qian}, \citenamefont
  {Sharma}, \citenamefont {Zhuang}, \citenamefont {Kalinin}, \citenamefont
  {Sumpter}, \citenamefont {Pilania}, \citenamefont {Acar}, \citenamefont
  {Mandal}, \citenamefont {Haule}, \citenamefont {Vanderbilt}, \citenamefont
  {Rabe},\ and\ \citenamefont {Tavazza}}]{db_jarvis}%
  \BibitemOpen
  \bibfield  {author} {\bibinfo {author} {\bibfnamefont {K.}~\bibnamefont
  {Choudhary}}, \bibinfo {author} {\bibfnamefont {K.~F.}\ \bibnamefont
  {Garrity}}, \bibinfo {author} {\bibfnamefont {A.~C.~E.}\ \bibnamefont
  {Reid}}, \bibinfo {author} {\bibfnamefont {B.}~\bibnamefont {DeCost}},
  \bibinfo {author} {\bibfnamefont {A.~J.}\ \bibnamefont {Biacchi}}, \bibinfo
  {author} {\bibfnamefont {A.~R.}\ \bibnamefont {Hight~Walker}}, \bibinfo
  {author} {\bibfnamefont {Z.}~\bibnamefont {Trautt}}, \bibinfo {author}
  {\bibfnamefont {J.}~\bibnamefont {Hattrick-Simpers}}, \bibinfo {author}
  {\bibfnamefont {A.~G.}\ \bibnamefont {Kusne}}, \bibinfo {author}
  {\bibfnamefont {A.}~\bibnamefont {Centrone}}, \bibinfo {author}
  {\bibfnamefont {A.}~\bibnamefont {Davydov}}, \bibinfo {author} {\bibfnamefont
  {J.}~\bibnamefont {Jiang}}, \bibinfo {author} {\bibfnamefont
  {R.}~\bibnamefont {Pachter}}, \bibinfo {author} {\bibfnamefont
  {G.}~\bibnamefont {Cheon}}, \bibinfo {author} {\bibfnamefont
  {E.}~\bibnamefont {Reed}}, \bibinfo {author} {\bibfnamefont {A.}~\bibnamefont
  {Agrawal}}, \bibinfo {author} {\bibfnamefont {X.}~\bibnamefont {Qian}},
  \bibinfo {author} {\bibfnamefont {V.}~\bibnamefont {Sharma}}, \bibinfo
  {author} {\bibfnamefont {H.}~\bibnamefont {Zhuang}}, \bibinfo {author}
  {\bibfnamefont {S.~V.}\ \bibnamefont {Kalinin}}, \bibinfo {author}
  {\bibfnamefont {B.~G.}\ \bibnamefont {Sumpter}}, \bibinfo {author}
  {\bibfnamefont {G.}~\bibnamefont {Pilania}}, \bibinfo {author} {\bibfnamefont
  {P.}~\bibnamefont {Acar}}, \bibinfo {author} {\bibfnamefont {S.}~\bibnamefont
  {Mandal}}, \bibinfo {author} {\bibfnamefont {K.}~\bibnamefont {Haule}},
  \bibinfo {author} {\bibfnamefont {D.}~\bibnamefont {Vanderbilt}}, \bibinfo
  {author} {\bibfnamefont {K.}~\bibnamefont {Rabe}},\ and\ \bibinfo {author}
  {\bibfnamefont {F.}~\bibnamefont {Tavazza}},\ }\href
  {https://doi.org/10.1038/s41524-020-00440-1} {\bibfield  {journal} {\bibinfo
  {journal} {npj Computational Materials}\ }\textbf {\bibinfo {volume} {6}},\
  \bibinfo {pages} {173} (\bibinfo {year} {2020})}\BibitemShut {NoStop}%
\bibitem [{\citenamefont {Choudhary}\ and\ \citenamefont
  {DeCost}(2021)}]{models_str_ALIGNN}%
  \BibitemOpen
  \bibfield  {author} {\bibinfo {author} {\bibfnamefont {K.}~\bibnamefont
  {Choudhary}}\ and\ \bibinfo {author} {\bibfnamefont {B.}~\bibnamefont
  {DeCost}},\ }\href {https://doi.org/10.1038/s41524-021-00650-1} {\bibfield
  {journal} {\bibinfo  {journal} {npj Computational Materials}\ }\textbf
  {\bibinfo {volume} {7}},\ \bibinfo {pages} {185} (\bibinfo {year}
  {2021})}\BibitemShut {NoStop}%
\bibitem [{\citenamefont {Jain}\ \emph {et~al.}(2013)\citenamefont {Jain},
  \citenamefont {Ong}, \citenamefont {Hautier}, \citenamefont {Chen},
  \citenamefont {Richards}, \citenamefont {Dacek}, \citenamefont {Cholia},
  \citenamefont {Gunter}, \citenamefont {Skinner}, \citenamefont {Ceder},\ and\
  \citenamefont {Persson}}]{db_mp}%
  \BibitemOpen
  \bibfield  {author} {\bibinfo {author} {\bibfnamefont {A.}~\bibnamefont
  {Jain}}, \bibinfo {author} {\bibfnamefont {S.~P.}\ \bibnamefont {Ong}},
  \bibinfo {author} {\bibfnamefont {G.}~\bibnamefont {Hautier}}, \bibinfo
  {author} {\bibfnamefont {W.}~\bibnamefont {Chen}}, \bibinfo {author}
  {\bibfnamefont {W.~D.}\ \bibnamefont {Richards}}, \bibinfo {author}
  {\bibfnamefont {S.}~\bibnamefont {Dacek}}, \bibinfo {author} {\bibfnamefont
  {S.}~\bibnamefont {Cholia}}, \bibinfo {author} {\bibfnamefont
  {D.}~\bibnamefont {Gunter}}, \bibinfo {author} {\bibfnamefont
  {D.}~\bibnamefont {Skinner}}, \bibinfo {author} {\bibfnamefont
  {G.}~\bibnamefont {Ceder}},\ and\ \bibinfo {author} {\bibfnamefont {K.~A.}\
  \bibnamefont {Persson}},\ }\href {https://doi.org/10.1063/1.4812323}
  {\bibfield  {journal} {\bibinfo  {journal} {APL Materials}\ }\textbf
  {\bibinfo {volume} {1}},\ \bibinfo {pages} {011002} (\bibinfo {year}
  {2013})},\ \Eprint {https://arxiv.org/abs/https://doi.org/10.1063/1.4812323}
  {https://doi.org/10.1063/1.4812323} \BibitemShut {NoStop}%
\bibitem [{\citenamefont {Davariashtiyani}\ and\ \citenamefont
  {Kadkhodaei}(2023)}]{davariashtiyani_2023_FE}%
  \BibitemOpen
  \bibfield  {author} {\bibinfo {author} {\bibfnamefont {A.}~\bibnamefont
  {Davariashtiyani}}\ and\ \bibinfo {author} {\bibfnamefont {S.}~\bibnamefont
  {Kadkhodaei}},\ }\bibfield  {journal} {\bibinfo  {journal} {PREPRINT}\ }\href
  {https://doi.org/10.21203/rs.3.rs-3044141/v1} {10.21203/rs.3.rs-3044141/v1}
  (\bibinfo {year} {2023})\BibitemShut {NoStop}%
\bibitem [{\citenamefont {Jiang}\ \emph {et~al.}(2021)\citenamefont {Jiang},
  \citenamefont {Chen}, \citenamefont {Chen}, \citenamefont {Li}, \citenamefont
  {Wei},\ and\ \citenamefont {Pan}}]{Jiang2021}%
  \BibitemOpen
  \bibfield  {author} {\bibinfo {author} {\bibfnamefont {Y.}~\bibnamefont
  {Jiang}}, \bibinfo {author} {\bibfnamefont {D.}~\bibnamefont {Chen}},
  \bibinfo {author} {\bibfnamefont {X.}~\bibnamefont {Chen}}, \bibinfo {author}
  {\bibfnamefont {T.}~\bibnamefont {Li}}, \bibinfo {author} {\bibfnamefont
  {G.-W.}\ \bibnamefont {Wei}},\ and\ \bibinfo {author} {\bibfnamefont
  {F.}~\bibnamefont {Pan}},\ }\href
  {https://doi.org/10.1038/s41524-021-00493-w} {\bibfield  {journal} {\bibinfo
  {journal} {npj Computational Materials}\ }\textbf {\bibinfo {volume} {7}},\
  \bibinfo {pages} {28} (\bibinfo {year} {2021})}\BibitemShut {NoStop}%
\bibitem [{\citenamefont {Jones}\ and\ \citenamefont
  {Stevanovi\ifmmode~\acute{c}\else
  \'{c}\fi{}}(2017)}]{Polymorphism_Jones2017}%
  \BibitemOpen
  \bibfield  {author} {\bibinfo {author} {\bibfnamefont {E.~B.}\ \bibnamefont
  {Jones}}\ and\ \bibinfo {author} {\bibfnamefont {V.}~\bibnamefont
  {Stevanovi\ifmmode~\acute{c}\else \'{c}\fi{}}},\ }\href
  {https://doi.org/10.1103/PhysRevB.96.184101} {\bibfield  {journal} {\bibinfo
  {journal} {Phys. Rev. B}\ }\textbf {\bibinfo {volume} {96}},\ \bibinfo
  {pages} {184101} (\bibinfo {year} {2017})}\BibitemShut {NoStop}%
\bibitem [{\citenamefont {Zhu}\ \emph {et~al.}(2023)\citenamefont {Zhu},
  \citenamefont {Tian}, \citenamefont {Ren}, \citenamefont {Li}, \citenamefont
  {Buonassisi},\ and\ \citenamefont {Hippalgaonkar}}]{syn_Zhu2023}%
  \BibitemOpen
  \bibfield  {author} {\bibinfo {author} {\bibfnamefont {R.}~\bibnamefont
  {Zhu}}, \bibinfo {author} {\bibfnamefont {S.~I.~P.}\ \bibnamefont {Tian}},
  \bibinfo {author} {\bibfnamefont {Z.}~\bibnamefont {Ren}}, \bibinfo {author}
  {\bibfnamefont {J.}~\bibnamefont {Li}}, \bibinfo {author} {\bibfnamefont
  {T.}~\bibnamefont {Buonassisi}},\ and\ \bibinfo {author} {\bibfnamefont
  {K.}~\bibnamefont {Hippalgaonkar}},\ }\href
  {https://doi.org/10.1021/acsomega.2c04856} {\bibfield  {journal} {\bibinfo
  {journal} {ACS Omega}\ }\textbf {\bibinfo {volume} {8}},\ \bibinfo {pages}
  {8210} (\bibinfo {year} {2023})}\BibitemShut {NoStop}%
\bibitem [{\citenamefont {Raccuglia}\ \emph {et~al.}(2016)\citenamefont
  {Raccuglia}, \citenamefont {Elbert}, \citenamefont {Adler}, \citenamefont
  {Falk}, \citenamefont {Wenny}, \citenamefont {Mollo}, \citenamefont {Zeller},
  \citenamefont {Friedler}, \citenamefont {Schrier},\ and\ \citenamefont
  {Norquist}}]{syn_Raccuglia2016_failed}%
  \BibitemOpen
  \bibfield  {author} {\bibinfo {author} {\bibfnamefont {P.}~\bibnamefont
  {Raccuglia}}, \bibinfo {author} {\bibfnamefont {K.~C.}\ \bibnamefont
  {Elbert}}, \bibinfo {author} {\bibfnamefont {P.~D.~F.}\ \bibnamefont
  {Adler}}, \bibinfo {author} {\bibfnamefont {C.}~\bibnamefont {Falk}},
  \bibinfo {author} {\bibfnamefont {M.~B.}\ \bibnamefont {Wenny}}, \bibinfo
  {author} {\bibfnamefont {A.}~\bibnamefont {Mollo}}, \bibinfo {author}
  {\bibfnamefont {M.}~\bibnamefont {Zeller}}, \bibinfo {author} {\bibfnamefont
  {S.~A.}\ \bibnamefont {Friedler}}, \bibinfo {author} {\bibfnamefont
  {J.}~\bibnamefont {Schrier}},\ and\ \bibinfo {author} {\bibfnamefont {A.~J.}\
  \bibnamefont {Norquist}},\ }\href {https://doi.org/10.1038/nature17439}
  {\bibfield  {journal} {\bibinfo  {journal} {Nature}\ }\textbf {\bibinfo
  {volume} {533}},\ \bibinfo {pages} {73} (\bibinfo {year} {2016})}\BibitemShut
  {NoStop}%
\bibitem [{\citenamefont {Kim}\ \emph {et~al.}(2017)\citenamefont {Kim},
  \citenamefont {Huang}, \citenamefont {Jegelka},\ and\ \citenamefont
  {Olivetti}}]{syn_kim_2017_params_routes}%
  \BibitemOpen
  \bibfield  {author} {\bibinfo {author} {\bibfnamefont {E.}~\bibnamefont
  {Kim}}, \bibinfo {author} {\bibfnamefont {K.}~\bibnamefont {Huang}}, \bibinfo
  {author} {\bibfnamefont {S.}~\bibnamefont {Jegelka}},\ and\ \bibinfo {author}
  {\bibfnamefont {E.}~\bibnamefont {Olivetti}},\ }\href
  {https://doi.org/10.1038/s41524-017-0055-6} {\bibfield  {journal} {\bibinfo
  {journal} {npj Computational Materials}\ }\textbf {\bibinfo {volume} {3}},\
  \bibinfo {pages} {53} (\bibinfo {year} {2017})}\BibitemShut {NoStop}%
\bibitem [{\citenamefont {Huo}\ \emph {et~al.}(2019)\citenamefont {Huo},
  \citenamefont {Rong}, \citenamefont {Kononova}, \citenamefont {Sun},
  \citenamefont {Botari}, \citenamefont {He}, \citenamefont {Tshitoyan},\ and\
  \citenamefont {Ceder}}]{syn_Huo2019_procedures}%
  \BibitemOpen
  \bibfield  {author} {\bibinfo {author} {\bibfnamefont {H.}~\bibnamefont
  {Huo}}, \bibinfo {author} {\bibfnamefont {Z.}~\bibnamefont {Rong}}, \bibinfo
  {author} {\bibfnamefont {O.}~\bibnamefont {Kononova}}, \bibinfo {author}
  {\bibfnamefont {W.}~\bibnamefont {Sun}}, \bibinfo {author} {\bibfnamefont
  {T.}~\bibnamefont {Botari}}, \bibinfo {author} {\bibfnamefont
  {T.}~\bibnamefont {He}}, \bibinfo {author} {\bibfnamefont {V.}~\bibnamefont
  {Tshitoyan}},\ and\ \bibinfo {author} {\bibfnamefont {G.}~\bibnamefont
  {Ceder}},\ }\href {https://doi.org/10.1038/s41524-019-0204-1} {\bibfield
  {journal} {\bibinfo  {journal} {npj Computational Materials}\ }\textbf
  {\bibinfo {volume} {5}},\ \bibinfo {pages} {62} (\bibinfo {year}
  {2019})}\BibitemShut {NoStop}%
\bibitem [{\citenamefont {Kononova}\ \emph {et~al.}(2019)\citenamefont
  {Kononova}, \citenamefont {Huo}, \citenamefont {He}, \citenamefont {Rong},
  \citenamefont {Botari}, \citenamefont {Sun}, \citenamefont {Tshitoyan},\ and\
  \citenamefont {Ceder}}]{syn_Kononova2019_recipes}%
  \BibitemOpen
  \bibfield  {author} {\bibinfo {author} {\bibfnamefont {O.}~\bibnamefont
  {Kononova}}, \bibinfo {author} {\bibfnamefont {H.}~\bibnamefont {Huo}},
  \bibinfo {author} {\bibfnamefont {T.}~\bibnamefont {He}}, \bibinfo {author}
  {\bibfnamefont {Z.}~\bibnamefont {Rong}}, \bibinfo {author} {\bibfnamefont
  {T.}~\bibnamefont {Botari}}, \bibinfo {author} {\bibfnamefont
  {W.}~\bibnamefont {Sun}}, \bibinfo {author} {\bibfnamefont {V.}~\bibnamefont
  {Tshitoyan}},\ and\ \bibinfo {author} {\bibfnamefont {G.}~\bibnamefont
  {Ceder}},\ }\href {https://doi.org/10.1038/s41597-019-0224-1} {\bibfield
  {journal} {\bibinfo  {journal} {Scientific Data}\ }\textbf {\bibinfo {volume}
  {6}},\ \bibinfo {pages} {203} (\bibinfo {year} {2019})}\BibitemShut {NoStop}%
\bibitem [{\citenamefont {Kim}\ \emph {et~al.}(2020)\citenamefont {Kim},
  \citenamefont {Jensen}, \citenamefont {van Grootel}, \citenamefont {Huang},
  \citenamefont {Staib}, \citenamefont {Mysore}, \citenamefont {Chang},
  \citenamefont {Strubell}, \citenamefont {McCallum}, \citenamefont {Jegelka},\
  and\ \citenamefont {Olivetti}}]{syn_kim_2020_precursor}%
  \BibitemOpen
  \bibfield  {author} {\bibinfo {author} {\bibfnamefont {E.}~\bibnamefont
  {Kim}}, \bibinfo {author} {\bibfnamefont {Z.}~\bibnamefont {Jensen}},
  \bibinfo {author} {\bibfnamefont {A.}~\bibnamefont {van Grootel}}, \bibinfo
  {author} {\bibfnamefont {K.}~\bibnamefont {Huang}}, \bibinfo {author}
  {\bibfnamefont {M.}~\bibnamefont {Staib}}, \bibinfo {author} {\bibfnamefont
  {S.}~\bibnamefont {Mysore}}, \bibinfo {author} {\bibfnamefont {H.-S.}\
  \bibnamefont {Chang}}, \bibinfo {author} {\bibfnamefont {E.}~\bibnamefont
  {Strubell}}, \bibinfo {author} {\bibfnamefont {A.}~\bibnamefont {McCallum}},
  \bibinfo {author} {\bibfnamefont {S.}~\bibnamefont {Jegelka}},\ and\ \bibinfo
  {author} {\bibfnamefont {E.}~\bibnamefont {Olivetti}},\ }\href
  {https://doi.org/10.1021/acs.jcim.9b00995} {\bibfield  {journal} {\bibinfo
  {journal} {J Chem Inf Model}\ }\textbf {\bibinfo {volume} {60}},\ \bibinfo
  {pages} {1194} (\bibinfo {year} {2020})}\BibitemShut {NoStop}%
\bibitem [{\citenamefont {Karpovich}\ \emph {et~al.}(2021)\citenamefont
  {Karpovich}, \citenamefont {Jensen}, \citenamefont {Venugopal},\ and\
  \citenamefont {Olivetti}}]{syn_karpovich2021inorganic_rout_condition}%
  \BibitemOpen
  \bibfield  {author} {\bibinfo {author} {\bibfnamefont {C.}~\bibnamefont
  {Karpovich}}, \bibinfo {author} {\bibfnamefont {Z.}~\bibnamefont {Jensen}},
  \bibinfo {author} {\bibfnamefont {V.}~\bibnamefont {Venugopal}},\ and\
  \bibinfo {author} {\bibfnamefont {E.}~\bibnamefont {Olivetti}},\ }\href@noop
  {} {\bibinfo {title} {Inorganic synthesis reaction condition prediction with
  generative machine learning}} (\bibinfo {year} {2021}),\ \Eprint
  {https://arxiv.org/abs/2112.09612} {arXiv:2112.09612 [cond-mat.mtrl-sci]}
  \BibitemShut {NoStop}%
\bibitem [{\citenamefont {Wang}\ \emph {et~al.}(2022)\citenamefont {Wang},
  \citenamefont {Kononova}, \citenamefont {Cruse}, \citenamefont {He},
  \citenamefont {Huo}, \citenamefont {Fei}, \citenamefont {Zeng}, \citenamefont
  {Sun}, \citenamefont {Cai}, \citenamefont {Sun},\ and\ \citenamefont
  {Ceder}}]{syn_Wang2022_procedures}%
  \BibitemOpen
  \bibfield  {author} {\bibinfo {author} {\bibfnamefont {Z.}~\bibnamefont
  {Wang}}, \bibinfo {author} {\bibfnamefont {O.}~\bibnamefont {Kononova}},
  \bibinfo {author} {\bibfnamefont {K.}~\bibnamefont {Cruse}}, \bibinfo
  {author} {\bibfnamefont {T.}~\bibnamefont {He}}, \bibinfo {author}
  {\bibfnamefont {H.}~\bibnamefont {Huo}}, \bibinfo {author} {\bibfnamefont
  {Y.}~\bibnamefont {Fei}}, \bibinfo {author} {\bibfnamefont {Y.}~\bibnamefont
  {Zeng}}, \bibinfo {author} {\bibfnamefont {Y.}~\bibnamefont {Sun}}, \bibinfo
  {author} {\bibfnamefont {Z.}~\bibnamefont {Cai}}, \bibinfo {author}
  {\bibfnamefont {W.}~\bibnamefont {Sun}},\ and\ \bibinfo {author}
  {\bibfnamefont {G.}~\bibnamefont {Ceder}},\ }\href
  {https://doi.org/10.1038/s41597-022-01317-2} {\bibfield  {journal} {\bibinfo
  {journal} {Scientific Data}\ }\textbf {\bibinfo {volume} {9}},\ \bibinfo
  {pages} {231} (\bibinfo {year} {2022})}\BibitemShut {NoStop}%
\bibitem [{\citenamefont {Huo}\ \emph {et~al.}(2022)\citenamefont {Huo},
  \citenamefont {Bartel}, \citenamefont {He}, \citenamefont {Trewartha},
  \citenamefont {Dunn}, \citenamefont {Ouyang}, \citenamefont {Jain},\ and\
  \citenamefont {Ceder}}]{syn_Huo_2022_condition}%
  \BibitemOpen
  \bibfield  {author} {\bibinfo {author} {\bibfnamefont {H.}~\bibnamefont
  {Huo}}, \bibinfo {author} {\bibfnamefont {C.}~\bibnamefont {Bartel}},
  \bibinfo {author} {\bibfnamefont {T.}~\bibnamefont {He}}, \bibinfo {author}
  {\bibfnamefont {A.}~\bibnamefont {Trewartha}}, \bibinfo {author}
  {\bibfnamefont {A.}~\bibnamefont {Dunn}}, \bibinfo {author} {\bibfnamefont
  {B.}~\bibnamefont {Ouyang}}, \bibinfo {author} {\bibfnamefont
  {A.}~\bibnamefont {Jain}},\ and\ \bibinfo {author} {\bibfnamefont
  {G.}~\bibnamefont {Ceder}},\ }\href
  {https://doi.org/10.1021/acs.chemmater.2c01293} {\bibfield  {journal}
  {\bibinfo  {journal} {Chemistry of Materials}\ }\textbf {\bibinfo {volume}
  {34}},\ \bibinfo {pages} {7323} (\bibinfo {year} {2022})},\ \bibinfo {note}
  {publisher Copyright: {\textcopyright} 2022 American Chemical Society. All
  rights reserved.}\BibitemShut {Stop}%
\bibitem [{\citenamefont {Karpovich}\ \emph {et~al.}(2023)\citenamefont
  {Karpovich}, \citenamefont {Pan}, \citenamefont {Jensen},\ and\ \citenamefont
  {Olivetti}}]{syn_Karpovich_2023_condition}%
  \BibitemOpen
  \bibfield  {author} {\bibinfo {author} {\bibfnamefont {C.}~\bibnamefont
  {Karpovich}}, \bibinfo {author} {\bibfnamefont {E.}~\bibnamefont {Pan}},
  \bibinfo {author} {\bibfnamefont {Z.}~\bibnamefont {Jensen}},\ and\ \bibinfo
  {author} {\bibfnamefont {E.}~\bibnamefont {Olivetti}},\ }\href
  {https://doi.org/10.1021/acs.chemmater.2c03010} {\bibfield  {journal}
  {\bibinfo  {journal} {Chemistry of Materials}\ }\textbf {\bibinfo {volume}
  {35}},\ \bibinfo {pages} {1062} (\bibinfo {year} {2023})}\BibitemShut
  {NoStop}%
\bibitem [{\citenamefont {McDermott}\ \emph {et~al.}(2023)\citenamefont
  {McDermott}, \citenamefont {McBride}, \citenamefont {Regier}, \citenamefont
  {Tran}, \citenamefont {Chen}, \citenamefont {Corrao}, \citenamefont
  {Gallant}, \citenamefont {Kamm}, \citenamefont {Bartel}, \citenamefont
  {Chapman}, \citenamefont {Khalifah}, \citenamefont {Ceder}, \citenamefont
  {Neilson},\ and\ \citenamefont {Persson}}]{syn_McDermott_2023_procedures}%
  \BibitemOpen
  \bibfield  {author} {\bibinfo {author} {\bibfnamefont {M.~J.}\ \bibnamefont
  {McDermott}}, \bibinfo {author} {\bibfnamefont {B.~C.}\ \bibnamefont
  {McBride}}, \bibinfo {author} {\bibfnamefont {C.~E.}\ \bibnamefont {Regier}},
  \bibinfo {author} {\bibfnamefont {G.~T.}\ \bibnamefont {Tran}}, \bibinfo
  {author} {\bibfnamefont {Y.}~\bibnamefont {Chen}}, \bibinfo {author}
  {\bibfnamefont {A.~A.}\ \bibnamefont {Corrao}}, \bibinfo {author}
  {\bibfnamefont {M.~C.}\ \bibnamefont {Gallant}}, \bibinfo {author}
  {\bibfnamefont {G.~E.}\ \bibnamefont {Kamm}}, \bibinfo {author}
  {\bibfnamefont {C.~J.}\ \bibnamefont {Bartel}}, \bibinfo {author}
  {\bibfnamefont {K.~W.}\ \bibnamefont {Chapman}}, \bibinfo {author}
  {\bibfnamefont {P.~G.}\ \bibnamefont {Khalifah}}, \bibinfo {author}
  {\bibfnamefont {G.}~\bibnamefont {Ceder}}, \bibinfo {author} {\bibfnamefont
  {J.~R.}\ \bibnamefont {Neilson}},\ and\ \bibinfo {author} {\bibfnamefont
  {K.~A.}\ \bibnamefont {Persson}},\ }\href
  {https://doi.org/10.1021/acscentsci.3c01051} {\bibfield  {journal} {\bibinfo
  {journal} {ACS Central Science}\ }\textbf {\bibinfo {volume} {9}},\ \bibinfo
  {pages} {1957} (\bibinfo {year} {2023})}\BibitemShut {NoStop}%
\bibitem [{\citenamefont {Aykol}\ \emph {et~al.}(2019)\citenamefont {Aykol},
  \citenamefont {Hegde}, \citenamefont {Hung}, \citenamefont {Suram},
  \citenamefont {Herring}, \citenamefont {Wolverton},\ and\ \citenamefont
  {Hummelsh{\o}j}}]{syn_Aykol2019_network}%
  \BibitemOpen
  \bibfield  {author} {\bibinfo {author} {\bibfnamefont {M.}~\bibnamefont
  {Aykol}}, \bibinfo {author} {\bibfnamefont {V.~I.}\ \bibnamefont {Hegde}},
  \bibinfo {author} {\bibfnamefont {L.}~\bibnamefont {Hung}}, \bibinfo {author}
  {\bibfnamefont {S.}~\bibnamefont {Suram}}, \bibinfo {author} {\bibfnamefont
  {P.}~\bibnamefont {Herring}}, \bibinfo {author} {\bibfnamefont
  {C.}~\bibnamefont {Wolverton}},\ and\ \bibinfo {author} {\bibfnamefont
  {J.~S.}\ \bibnamefont {Hummelsh{\o}j}},\ }\href
  {https://doi.org/10.1038/s41467-019-10030-5} {\bibfield  {journal} {\bibinfo
  {journal} {Nature Communications}\ }\textbf {\bibinfo {volume} {10}},\
  \bibinfo {pages} {2018} (\bibinfo {year} {2019})}\BibitemShut {NoStop}%
\bibitem [{\citenamefont {Aykol}\ \emph {et~al.}(2021)\citenamefont {Aykol},
  \citenamefont {Montoya},\ and\ \citenamefont
  {Hummelsh{\o}j}}]{syn_Aykol2021_pareto}%
  \BibitemOpen
  \bibfield  {author} {\bibinfo {author} {\bibfnamefont {M.}~\bibnamefont
  {Aykol}}, \bibinfo {author} {\bibfnamefont {J.~H.}\ \bibnamefont {Montoya}},\
  and\ \bibinfo {author} {\bibfnamefont {J.}~\bibnamefont {Hummelsh{\o}j}},\
  }\href {https://doi.org/10.1021/jacs.1c04888} {\bibfield  {journal} {\bibinfo
   {journal} {Journal of the American Chemical Society}\ }\textbf {\bibinfo
  {volume} {143}},\ \bibinfo {pages} {9244} (\bibinfo {year}
  {2021})}\BibitemShut {NoStop}%
\bibitem [{\citenamefont {McDermott}\ \emph {et~al.}(2021)\citenamefont
  {McDermott}, \citenamefont {Dwaraknath},\ and\ \citenamefont
  {Persson}}]{syn_McDermott2021_network}%
  \BibitemOpen
  \bibfield  {author} {\bibinfo {author} {\bibfnamefont {M.~J.}\ \bibnamefont
  {McDermott}}, \bibinfo {author} {\bibfnamefont {S.~S.}\ \bibnamefont
  {Dwaraknath}},\ and\ \bibinfo {author} {\bibfnamefont {K.~A.}\ \bibnamefont
  {Persson}},\ }\href {https://doi.org/10.1038/s41467-021-23339-x} {\bibfield
  {journal} {\bibinfo  {journal} {Nature Communications}\ }\textbf {\bibinfo
  {volume} {12}},\ \bibinfo {pages} {3097} (\bibinfo {year}
  {2021})}\BibitemShut {NoStop}%
\bibitem [{\citenamefont {Gra{\v{z}}ulis}\ \emph {et~al.}(2009)\citenamefont
  {Gra{\v{z}}ulis}, \citenamefont {Chateigner}, \citenamefont {Downs},
  \citenamefont {Yokochi}, \citenamefont {Quir{\'{o}}s}, \citenamefont
  {Lutterotti}, \citenamefont {Manakova}, \citenamefont {Butkus}, \citenamefont
  {Moeck},\ and\ \citenamefont {Le~Bail}}]{Grazulis2009}%
  \BibitemOpen
  \bibfield  {author} {\bibinfo {author} {\bibfnamefont {S.}~\bibnamefont
  {Gra{\v{z}}ulis}}, \bibinfo {author} {\bibfnamefont {D.}~\bibnamefont
  {Chateigner}}, \bibinfo {author} {\bibfnamefont {R.~T.}\ \bibnamefont
  {Downs}}, \bibinfo {author} {\bibfnamefont {A.~F.~T.}\ \bibnamefont
  {Yokochi}}, \bibinfo {author} {\bibfnamefont {M.}~\bibnamefont
  {Quir{\'{o}}s}}, \bibinfo {author} {\bibfnamefont {L.}~\bibnamefont
  {Lutterotti}}, \bibinfo {author} {\bibfnamefont {E.}~\bibnamefont
  {Manakova}}, \bibinfo {author} {\bibfnamefont {J.}~\bibnamefont {Butkus}},
  \bibinfo {author} {\bibfnamefont {P.}~\bibnamefont {Moeck}},\ and\ \bibinfo
  {author} {\bibfnamefont {A.}~\bibnamefont {Le~Bail}},\ }\href
  {https://doi.org/10.1107/S0021889809016690} {\bibfield  {journal} {\bibinfo
  {journal} {Journal of Applied Crystallography}\ }\textbf {\bibinfo {volume}
  {42}},\ \bibinfo {pages} {726} (\bibinfo {year} {2009})}\BibitemShut
  {NoStop}%
\bibitem [{\citenamefont {Su}\ \emph {et~al.}(2017)\citenamefont {Su},
  \citenamefont {Lv}, \citenamefont {Li}, \citenamefont {Wang}, \citenamefont
  {Zhang}, \citenamefont {Wang},\ and\ \citenamefont {Ma}}]{db_cspd}%
  \BibitemOpen
  \bibfield  {author} {\bibinfo {author} {\bibfnamefont {C.}~\bibnamefont
  {Su}}, \bibinfo {author} {\bibfnamefont {J.}~\bibnamefont {Lv}}, \bibinfo
  {author} {\bibfnamefont {Q.}~\bibnamefont {Li}}, \bibinfo {author}
  {\bibfnamefont {H.}~\bibnamefont {Wang}}, \bibinfo {author} {\bibfnamefont
  {L.}~\bibnamefont {Zhang}}, \bibinfo {author} {\bibfnamefont
  {Y.}~\bibnamefont {Wang}},\ and\ \bibinfo {author} {\bibfnamefont
  {Y.}~\bibnamefont {Ma}},\ }\href {https://doi.org/10.1088/1361-648X/aa63cd}
  {\bibfield  {journal} {\bibinfo  {journal} {Journal of Physics: Condensed
  Matter}\ }\textbf {\bibinfo {volume} {29}},\ \bibinfo {pages} {165901}
  (\bibinfo {year} {2017})}\BibitemShut {NoStop}%
\bibitem [{\citenamefont {Zagorac}\ \emph {et~al.}(2019)\citenamefont
  {Zagorac}, \citenamefont {M{\"{u}}ller}, \citenamefont {Ruehl}, \citenamefont
  {Zagorac},\ and\ \citenamefont {Rehme}}]{db_icsd}%
  \BibitemOpen
  \bibfield  {author} {\bibinfo {author} {\bibfnamefont {D.}~\bibnamefont
  {Zagorac}}, \bibinfo {author} {\bibfnamefont {H.}~\bibnamefont
  {M{\"{u}}ller}}, \bibinfo {author} {\bibfnamefont {S.}~\bibnamefont {Ruehl}},
  \bibinfo {author} {\bibfnamefont {J.}~\bibnamefont {Zagorac}},\ and\ \bibinfo
  {author} {\bibfnamefont {S.}~\bibnamefont {Rehme}},\ }\href
  {https://doi.org/10.1107/S160057671900997X} {\bibfield  {journal} {\bibinfo
  {journal} {Journal of Applied Crystallography}\ }\textbf {\bibinfo {volume}
  {52}},\ \bibinfo {pages} {918} (\bibinfo {year} {2019})}\BibitemShut
  {NoStop}%
\bibitem [{\citenamefont {Quir{\'{o}}s}\ \emph {et~al.}(2018)\citenamefont
  {Quir{\'{o}}s}, \citenamefont {Gra{\v{z}}ulis}, \citenamefont
  {Girdzijauskait{\.{e}}}, \citenamefont {Merkys},\ and\ \citenamefont
  {Vaitkus}}]{Quiros2018}%
  \BibitemOpen
  \bibfield  {author} {\bibinfo {author} {\bibfnamefont {M.}~\bibnamefont
  {Quir{\'{o}}s}}, \bibinfo {author} {\bibfnamefont {S.}~\bibnamefont
  {Gra{\v{z}}ulis}}, \bibinfo {author} {\bibfnamefont {S.}~\bibnamefont
  {Girdzijauskait{\.{e}}}}, \bibinfo {author} {\bibfnamefont {A.}~\bibnamefont
  {Merkys}},\ and\ \bibinfo {author} {\bibfnamefont {A.}~\bibnamefont
  {Vaitkus}},\ }\bibfield  {journal} {\bibinfo  {journal} {Journal of
  Cheminformatics}\ }\textbf {\bibinfo {volume} {10}},\ \href
  {https://doi.org/10.1186/s13321-018-0279-6} {10.1186/s13321-018-0279-6}
  (\bibinfo {year} {2018})\BibitemShut {NoStop}%
\bibitem [{\citenamefont {Merkys}\ \emph {et~al.}(2016)\citenamefont {Merkys},
  \citenamefont {Vaitkus}, \citenamefont {Butkus}, \citenamefont
  {Okulič-Kazarinas}, \citenamefont {Kairys},\ and\ \citenamefont
  {Gražulis}}]{Merkys2016}%
  \BibitemOpen
  \bibfield  {author} {\bibinfo {author} {\bibfnamefont {A.}~\bibnamefont
  {Merkys}}, \bibinfo {author} {\bibfnamefont {A.}~\bibnamefont {Vaitkus}},
  \bibinfo {author} {\bibfnamefont {J.}~\bibnamefont {Butkus}}, \bibinfo
  {author} {\bibfnamefont {M.}~\bibnamefont {Okulič-Kazarinas}}, \bibinfo
  {author} {\bibfnamefont {V.}~\bibnamefont {Kairys}},\ and\ \bibinfo {author}
  {\bibfnamefont {S.}~\bibnamefont {Gražulis}},\ }\bibfield  {journal}
  {\bibinfo  {journal} {Journal of Applied Crystallography}\ }\textbf {\bibinfo
  {volume} {49}},\ \href {https://doi.org/10.1107/S1600576715022396}
  {10.1107/S1600576715022396} (\bibinfo {year} {2016})\BibitemShut {NoStop}%
\bibitem [{\citenamefont {Gražulis}\ \emph {et~al.}(2015)\citenamefont
  {Gražulis}, \citenamefont {Merkys}, \citenamefont {Vaitkus},\ and\
  \citenamefont {Okulič-Kazarinas}}]{Grazulis2015}%
  \BibitemOpen
  \bibfield  {author} {\bibinfo {author} {\bibfnamefont {S.}~\bibnamefont
  {Gražulis}}, \bibinfo {author} {\bibfnamefont {A.}~\bibnamefont {Merkys}},
  \bibinfo {author} {\bibfnamefont {A.}~\bibnamefont {Vaitkus}},\ and\ \bibinfo
  {author} {\bibfnamefont {M.}~\bibnamefont {Okulič-Kazarinas}},\ }\href
  {https://doi.org/10.1107/S1600576714025904} {\bibfield  {journal} {\bibinfo
  {journal} {Journal of Applied Crystallography}\ }\textbf {\bibinfo {volume}
  {48}},\ \bibinfo {pages} {85} (\bibinfo {year} {2015})}\BibitemShut {NoStop}%
\bibitem [{\citenamefont {Gražulis}\ \emph {et~al.}(2012)\citenamefont
  {Gražulis}, \citenamefont {Daškevič}, \citenamefont {Merkys},
  \citenamefont {Chateigner}, \citenamefont {Lutterotti}, \citenamefont
  {Quirós}, \citenamefont {Serebryanaya}, \citenamefont {Moeck}, \citenamefont
  {Downs},\ and\ \citenamefont {Le~Bail}}]{Grazulis2012}%
  \BibitemOpen
  \bibfield  {author} {\bibinfo {author} {\bibfnamefont {S.}~\bibnamefont
  {Gražulis}}, \bibinfo {author} {\bibfnamefont {A.}~\bibnamefont
  {Daškevič}}, \bibinfo {author} {\bibfnamefont {A.}~\bibnamefont {Merkys}},
  \bibinfo {author} {\bibfnamefont {D.}~\bibnamefont {Chateigner}}, \bibinfo
  {author} {\bibfnamefont {L.}~\bibnamefont {Lutterotti}}, \bibinfo {author}
  {\bibfnamefont {M.}~\bibnamefont {Quirós}}, \bibinfo {author} {\bibfnamefont
  {N.~R.}\ \bibnamefont {Serebryanaya}}, \bibinfo {author} {\bibfnamefont
  {P.}~\bibnamefont {Moeck}}, \bibinfo {author} {\bibfnamefont {R.~T.}\
  \bibnamefont {Downs}},\ and\ \bibinfo {author} {\bibfnamefont
  {A.}~\bibnamefont {Le~Bail}},\ }\href {https://doi.org/10.1093/nar/gkr900}
  {\bibfield  {journal} {\bibinfo  {journal} {Nucleic Acids Research}\ }\textbf
  {\bibinfo {volume} {40}},\ \bibinfo {pages} {D420} (\bibinfo {year}
  {2012})},\ \Eprint
  {https://arxiv.org/abs/http://nar.oxfordjournals.org/content/40/D1/D420.full.pdf+html}
  {http://nar.oxfordjournals.org/content/40/D1/D420.full.pdf+html} \BibitemShut
  {NoStop}%
\bibitem [{\citenamefont {Downs}\ and\ \citenamefont
  {Hall-Wallace}(2003)}]{Downs2003}%
  \BibitemOpen
  \bibfield  {author} {\bibinfo {author} {\bibfnamefont {R.~T.}\ \bibnamefont
  {Downs}}\ and\ \bibinfo {author} {\bibfnamefont {M.}~\bibnamefont
  {Hall-Wallace}},\ }\href@noop {} {\bibfield  {journal} {\bibinfo  {journal}
  {American Mineralogist}\ }\textbf {\bibinfo {volume} {88}},\ \bibinfo {pages}
  {247} (\bibinfo {year} {2003})}\BibitemShut {NoStop}%
\bibitem [{\citenamefont {Vaitkus}\ \emph {et~al.}(2023)\citenamefont
  {Vaitkus}, \citenamefont {Merkys}, \citenamefont {Sander}, \citenamefont
  {Quir{\'o}s}, \citenamefont {Thiessen}, \citenamefont {Bolton},\ and\
  \citenamefont {Gra{\v{z}}ulis}}]{db_cod}%
  \BibitemOpen
  \bibfield  {author} {\bibinfo {author} {\bibfnamefont {A.}~\bibnamefont
  {Vaitkus}}, \bibinfo {author} {\bibfnamefont {A.}~\bibnamefont {Merkys}},
  \bibinfo {author} {\bibfnamefont {T.}~\bibnamefont {Sander}}, \bibinfo
  {author} {\bibfnamefont {M.}~\bibnamefont {Quir{\'o}s}}, \bibinfo {author}
  {\bibfnamefont {P.~A.}\ \bibnamefont {Thiessen}}, \bibinfo {author}
  {\bibfnamefont {E.~E.}\ \bibnamefont {Bolton}},\ and\ \bibinfo {author}
  {\bibfnamefont {S.}~\bibnamefont {Gra{\v{z}}ulis}},\ }\href
  {https://doi.org/10.1186/s13321-023-00780-2} {\bibfield  {journal} {\bibinfo
  {journal} {Journal of Cheminformatics}\ }\textbf {\bibinfo {volume} {15}},\
  \bibinfo {pages} {123} (\bibinfo {year} {2023})}\BibitemShut {NoStop}%
\bibitem [{\citenamefont {Su}(2018)}]{CSPDGithub}%
  \BibitemOpen
  \bibfield  {author} {\bibinfo {author} {\bibfnamefont {C.}~\bibnamefont
  {Su}},\ }\href@noop {} {\bibinfo {title} {Atomic structure generator}},\
  \bibinfo {howpublished}
  {\url{https://github.com/SUNCAT-Center/AtomicStructureGenerator.git}}
  (\bibinfo {year} {2018})\BibitemShut {NoStop}%
\bibitem [{\citenamefont {Antunes}\ \emph {et~al.}(2024)\citenamefont
  {Antunes}, \citenamefont {Butler},\ and\ \citenamefont
  {Grau-Crespo}}]{antunes2024crystal}%
  \BibitemOpen
  \bibfield  {author} {\bibinfo {author} {\bibfnamefont {L.~M.}\ \bibnamefont
  {Antunes}}, \bibinfo {author} {\bibfnamefont {K.~T.}\ \bibnamefont
  {Butler}},\ and\ \bibinfo {author} {\bibfnamefont {R.}~\bibnamefont
  {Grau-Crespo}},\ }\href@noop {} {\bibinfo {title} {Crystal structure
  generation with autoregressive large language modeling}} (\bibinfo {year}
  {2024}),\ \Eprint {https://arxiv.org/abs/2307.04340} {arXiv:2307.04340
  [cond-mat.mtrl-sci]} \BibitemShut {NoStop}%
\end{thebibliography}
%

\end{document}